\documentclass[structabstract]{aa}
\usepackage{txfonts}
\usepackage{graphicx}
\usepackage{natbib}
\usepackage{lscape}
\usepackage{wasysym}
\bibpunct{(}{)}{;}{a}{}{,} 

\begin{document}

\title{Planetary detection limits taking into account stellar noise
\thanks{Based on observations collected at the La Silla Parana Observatory,
ESO (Chile), with the HARPS spectrograph at the 3.6-m telescope.}}

\subtitle{I. Observational strategies to reduce stellar oscillation and granulation effects}

\author{
  X. Dumusque\inst{1,2}\and
  S. Udry\inst{2}\and
  C. Lovis\inst{2}\and
  N.C. Santos\inst{1,3}\and
  M. J. P. F. G. Monteiro\inst{1,3}
  }

\institute{
    Centro de Astrof{\'\i}sica, Universidade do Porto, Rua das Estrelas, 4150-762 Porto, Portugal \and
    Observatoire de Gen\`eve, Universit\'e de Genve, 51 ch. des Maillettes, 1290 Sauverny, Switzerland \and
    Departamento de F{\'\i}sica e Astronomia, Faculdade de Ci\^encias da Universidade do Porto, Portugal
}

\date{Received XXX; accepted XXX}

\abstract
{Stellar noise produced by oscillations, granulation phenomena (granulation, mesogranulation and supergranulation) and activity affects radial velocity measurements. The signature of this noise in radial velocity is small, around the meter-per-second, but already too much for the detection of Earth mass planets in habitable zones.} 
{In this paper, we address the important role played by observational strategies in averaging out the radial velocity signature of stellar noise. We also derive the planetary mass detection limits expected in presence of stellar noise.}
{We start with HARPS asteroseismology measurements for 4 stars ($\beta$\,Hyi, $\alpha$\,Cen\,A, $\mu$\,Ara and $\tau$\,Ceti) available in the ESO archive plus very precise measurements of $\alpha$\,Cen\,B. This sample covers different spectral types, from G2 to K1 and different evolutionary stage, from subgiant to dwarf stars. Since the span of our data ranges between 5 to 8 days, only stellar noise sources with a time scale smaller than this time span will be extracted from these observations. Therefore, we will have access to oscillation modes and granulation phenomena, without important contribution of activity noise which is present at larger time scales. For those 5 stars, we generate synthetic radial velocity measurements after fitting corresponding models of stellar noise in Fourier space. These measurements allows us to study the radial velocity variation due to stellar noise for different observational strategies as well as the corresponding planetary mass detection limits.}
{Applying 3 measurements per night of 10 minutes exposure each, 2 hours apart, seems to average out most efficiently the stellar noise considered. For quiet K1V stars as $\alpha$\,Cen\,B, such a strategy allows us to detect planets of $\sim$\,3 times the mass of Earth with an orbital period of 200 days, corresponding to the habitable zone of the star. Our simulations moreover suggest that planets smaller than typically 5\,M$_{\oplus}$ can be detected with HARPS over a wide range of separations around most non-active solar type dwarfs. Since activity is not yet included in our simulation, these detection limits correspond to a case, which exist, where the host star has few magnetic features. In this case stellar noise is dominated by oscillation modes and granulation phenomena. For our star sample, a trend between spectral type and surface gravity and the level of radial velocity variation is also emphasized by our simulations.} 
{}

\keywords{stars: individual:  $\beta$\,Hyi --
	     stars: individual: $\mu$\,Ara--
             stars: individual: $\alpha$\,Cen\,A -- 
             stars: individual: $\tau$\,Ceti -- 
             stars: individual: $\alpha$\,Cen\,B -- 
             stars:planetary systems --
  	    stars: oscilation --
	    techniques: radial velocities
	    }

\authorrunning{Dumusque et al.}
\titlerunning{Stellar noise and planetary detection}
\maketitle

\section{Introduction}

The majority of planets discovered by Doppler spectroscopy are gaseous giants similar to Jupiter, with masses of a hundred times the mass of the Earth\footnote{see The Extrasolar Planets Encyclopaedia, http://exoplanet.eu}(M$_{\oplus}$). However, since approximately 5 years, planets from 2 to 10 M$_{\oplus}$, have been detected \citep[e.g.][]{Mayor-2009a, Mayor-2009b, Udry-2007b}. This has become possible thanks to the stability and precision of new generation high-resolution spectrographs, and because of dedicated observational strategies allowing to average out perturbations coming from stellar oscillations \citep[][]{Santos-2004a}. 

Pressure waves (p-modes) propagate at the surface of solar-type stars leading to a dilatation and contraction of external envelopes over time scales of a few minutes \citep[5 to 15 minutes for the Sun;][]{Schrijver-2000,Broomhall-2009}. The radial-velocity signature of these modes is typically varying between 10 and 400\,cm\,s$^{-1}$, depending on the star type and evolutionary stage \citep[][]{Schrijver-2000}. The amplitude and period of oscillation modes increases with mass along the main sequence. Theory predicts that frequencies of p-modes rise up with the square root of the star mean density and that their amplitudes are proportional to the luminosity over mass ratio \citep[][]{Christensen-Dalsgaard-2004}. The most precise spectrograph nowadays, HARPS, can reach a radial-velocity precision better than the meter-per-second in a short exposure time for bright stars \citep[typically 1\,m\,s$^{-1}$ in 1 minute for a $V=7.5$ K dwarf,][]{Pepe-2005}. These acoustic modes are then directly observable and can mask an eventual smallest planet signature. 

The different phenomena of granulation\footnote{hereafter, the term \emph{granulation phenomena} will be used to describe all type of convective motion (granulation, mesogranulation and supergranulation), whereas \emph{granulation} will just be used for the smaller time scale convective motion.} (granulation, mesogranulation, and supergranulation), due to the convective nature of solar type stars, also affect radial velocity (RV) measurements. These convective phenomena can be found over all the stellar surface, except in active regions where convection is greatly reduced \citep[e.g.][]{Dravins-1982,Livingston-1982,Brandt-1990,Gray-1992,Meunier-2010}. The amplitudes of granulation phenomena are similar to the ones observed for pressure modes \citep[e.g.][]{Schrijver-2000,Kjeldsen-2005}. Granulation corresponds to a small convective pattern with a lifetime shorter than 25 minutes, and a diameter smaller than 2 \,Mm \citep[][]{Title-1989,Del_Moro-2004b}. At much larger scale, we can find supergranulation. This phenomenon, linked to very large convective patterns, 15 to 40 Mm, can have a lifetime up to 33 hours in the Sun \citep[][]{Del_Moro-2004a}. Mesogranulation represent a convective phenomenon which can be located between the granulation and supergranulation, in terms of size and lifetime \citep[][]{Harvey-1984,Palle-1995,Schrijver-2000}. At the moment, the HARPS-GTO strategy, for the very high precision programs, uses long exposure times (15 minutes), in order to reduce the effects produced by stellar oscillation on RV measurements. However, this does not reduce low frequency noises coming from granulation phenomena.

On the time scale of tenths of days, another type of noise, due to the presence of activity related spots and plages, may perturb precise RV measurements. The presence of spots and plages on the surface of the star will break the flux balance between the red-shifted and the blue-shifted halves of the star. As the star rotates, a spot, or a plage,  moves across the stellar disk and produces an apparent Doppler shift \citep[][]{Saar-1997,Queloz-2001,Huelamo-2008,Lagrange-2010}. This modulation can be hard to disentangle from the Doppler modulation caused by the gravitational pull of a planet. On the Sun, taken as a proxi for G stars, the amplitude of this perturbation can reach 40\,cm\,s$^{-1}$ in high activity phase \citep[][]{Meunier-2010}. Spots are cooler than the solar mean surface temperature and plages hotter. Since active regions contain both, the noise induced by spots and plages will usually compensate, but not entirely since the surface ratio between spots and plages varies \citep[e.g.][]{Chapman-2001}. According to \citet{Meunier-2010}, the major effect of activity is not due to the break of the flux balance but to the inhibition of convection in active regions. They found that the corresponding noise could vary between 40 cm\,s$^{-1}$ at minimum activity and 140 cm\,s$^{-1}$ at maximum.

In this paper, we present simulations exploring new measurement strategies allowing to average granulation phenomena and oscillation modes at the same time. The detection limits for different strategies are derived, pointing out in particular a strategy that efficiently optimizes detection ability and realistic observational cost. Our simulations also emphasize that for dwarf stars, the level of RV variation, mainly due to stellar noise, is correlated to the spectral type and/or the surface gravity, log g.

We note that in this study we will just focus on understanding the stellar noise influencing in time scales smaller than the span of our data, 8 days. On such time scales, only the influence of oscillation and granulation phenomena can be fully characterized. Longer term activity signals, typically modulated at timescales of $\sim$30\,days (the usual rotation period of solar type stars) are not fully accessible. However, noise coming from activity related phenomena are still present in our data. The obtained results will thus only be valid for stars without significant activity phenomena. The effect of plages and spots will be estimated in a forthcoming paper.

\section{Data selection}

In order to characterize as well as possible the RV noise intrinsic to Sun-like stars, we wanted the most precise asteroseismology measurements (continuous follow up of the star with a high sampling rate). Another requirement was to use only data from one instrument, to avoid any instrumental discrepancies. We thus choose to use only HARPS measurements, since this spectrograph can reach a RV precision better than the meter per second, on the long term \citep[][]{Pepe-2005}, and even a better precision, at the level of a few 10s of cm\,s$^{-1}$, over one night \citep[][]{Bouchy-2005c}. Four stars with adequate observations are available in the HARPS ESO archive: $\beta$\,Hyi (\object{HD\,2151}), $\mu$\,Ara (\object{HD\,160691}), $\alpha$\,Cen\,A (\object{HD\,128620}) and $\tau$\,Ceti (\object{HD\,10700}). The measurements of $\alpha$\,Cen\,B (\object{HD\,128621}) have been provided by Pepe, F. (private communication) and do not correspond to continuous measurements. This star has been observed for 8 consecutive nights, though only 3 blocks of 20 minutes (about 27 measurements of 13 seconds) were done during each night. We will see afterwards that this is sufficient to characterize all type of considered noises. Depending on the star, the total span of the measurements varies from 5 to 8 nights. Table \ref{tab:1} gives more informations about the stellar parameters. Our sample contains 3 main sequence (MS) stars, 1 sub-giant and 1 star, $\mu$\,Ara, which is between the 2 regimes \citep[][]{Soriano-2010}. This will allows us to study a possible correlation between the level of RV variation and spectral type and/or evolution.
\begin{table*}[htdp] 
\begin{center}
\caption{Properties of our sample stars. For each star, the dead time due to CCD readout is 31\,s ((a) \citet{Sousa-2008}, (b) \citet{Bedding-2007b}, (c) \citet{Bazot-2007}, (d) \citet{Bouchy-2005c}, (e) \citet{Teixeira-2009}). ST is for spectral type and T$_{exp}$ for exposure time.}  \label{tab:1}
\begin{tabular}{ccccccccccc}
\hline
\hline Star & M$_V$ & Mass [$M_{\odot}]^{(a)}$ & $T_{eff}$ [K]$^{(a)}$ & $[Fe/H]^{(a)}$ & log g$^{(a)}$ & log(R'$_{HK}$)& ST & \# meas. & \# nights/span [days] & T$_{exp}$ [s] \\
\hline
$\beta$\,Hyi & 2.8 & 1.10 & 5872 & -0.11 & 3.95 & - & G2lV & 2766$^{(b)}$ & 6/6 & 40 to 50$^{(b)}$\\
$\mu$\,Ara & 5.2 & 1.08 & 5780 & 0.30 & 4.27 & -5.20 & G3IV-V & 2104$^{(d)}$ & 8/8 & 100$^{(d)}$\\
$\alpha$\,Cen\,A & 0.0 & 1.09 & 5623 & 0.22 & 4.22 & -5.07 & G2V & 4959$^{(c)}$ & 5/5 & 2 to 10$^{(c)}$\\
$\tau$\,Ceti & 3.5 & 0.63 & 5310 & -0.52 & 4.44 & -4.93 & G8V & 1962$^{(e)}$ & 5/6 & 23 to 40$^{(e)}$\\
$\alpha$\,Cen\,B & 1.3 & 0.91 & 5248 & 0.26 & 4.55 & -4.98 & K1V & 783 & 8/8 & 13\\
\hline
\end{tabular} 
\end{center}
\end{table*}

Exposure times are always shorter than 100\,s. Due to CCD readout, a dead time of $\sim$30\,s exists between each exposure. Such a strategy defines precisely the structure of oscillation modes, since they have a typical time scale above 5 minutes. The continuous follow up of the star during 5 to 8 days is essential in order to characterize convective motions (granulation phenomena), which have time scales up to $\sim$30 hours. In the case of $\alpha$\,Cen\,B, the 20 minutes of continuous measurements allow us to characterize granulation, since it has typical time scales smaller than 25 minutes. The mesogranulation perturbations can be analyzed thanks to the repetition, 3 times a night, of the continuous measurements. And finally, the 8 consecutive nights allow us to address supergranulation. We will see in Sec. \ref{subsect:2.1}, that for the case of $\alpha$\,Cen\,B, all types of noise can be distinguish clearly in the power spectrum, as it is the case for the other stars with continuous measurements.

In our sample, $\mu$\,Ara is the only star known to have planetary companions. We clearly see in the RVs the effect of the small planet in the system, which we suppress taking the orbital parameters given in \citet{Pepe-2006}.

\section{Generating synthetic RV measurements}\label{sect:2}

In this paper, we study different observational strategies to average out as much as possible the effect of stellar noise on RV measurements. The main goal is to find a strategy capable of detecting very small mass planets in habitable regions. To be able to search for such long period planets, observational strategies must be efficient on several years. Since the asteroseismology data of the selected stars span only over a maximum of 8 days, we have to create longer synthetic RV data, containing the same structure and level of noise. To generate these synthetic data, we first calculate the velocity power spectrum density (VPSD) using the periodogram technique. We then fit the obtained spectrum with a noise model and go back to the RV space using the property that the periodogram is equivalent to fitting series of sine waves \citep[e.g.][]{Scargle-1982,Zechmeister-2009}. The transformation from RVs to the VPSD, as well as the inverse one, does not loose any information, so the synthetic RV sets should present the same noise structure as the real data we started from. We have however to keep in mind that even if we have synthetic RV data over more days then the real measurements, we do not  take into account, in our model, perturbing effects with time scales longer than the real observation time. Therefore activity noise is not fully included.


\subsection{Velocity power spectrum density} \label{subsect:2.0}

The first step is to calculate the velocity power spectrum of the asteroseismology RV measurements. Since the measurements are not evenly spaced in time, we can not use a standard FFT algorithm. We thus use a weighted Lomb-Scargle periodogram \citep[][]{Lomb-1976,Scargle-1982,Zechmeister-2009} to calculate the power spectrum of the HARPS RV measurements. The minimum frequency of the spectrum is set to 1/T, where T is the total span of the observing run and the maximum frequency is set to the Nyquist frequency for the median sampling. To obtain the VPSD, which is independent of the observing window, we multiply the power by the effective length of the observing run, calculated as the inverse of the area under the spectral window \citep[in power, see][]{Kjeldsen-2005}.

\begin{figure}[!]
\begin{center}
\resizebox{\hsize}{!}{\includegraphics{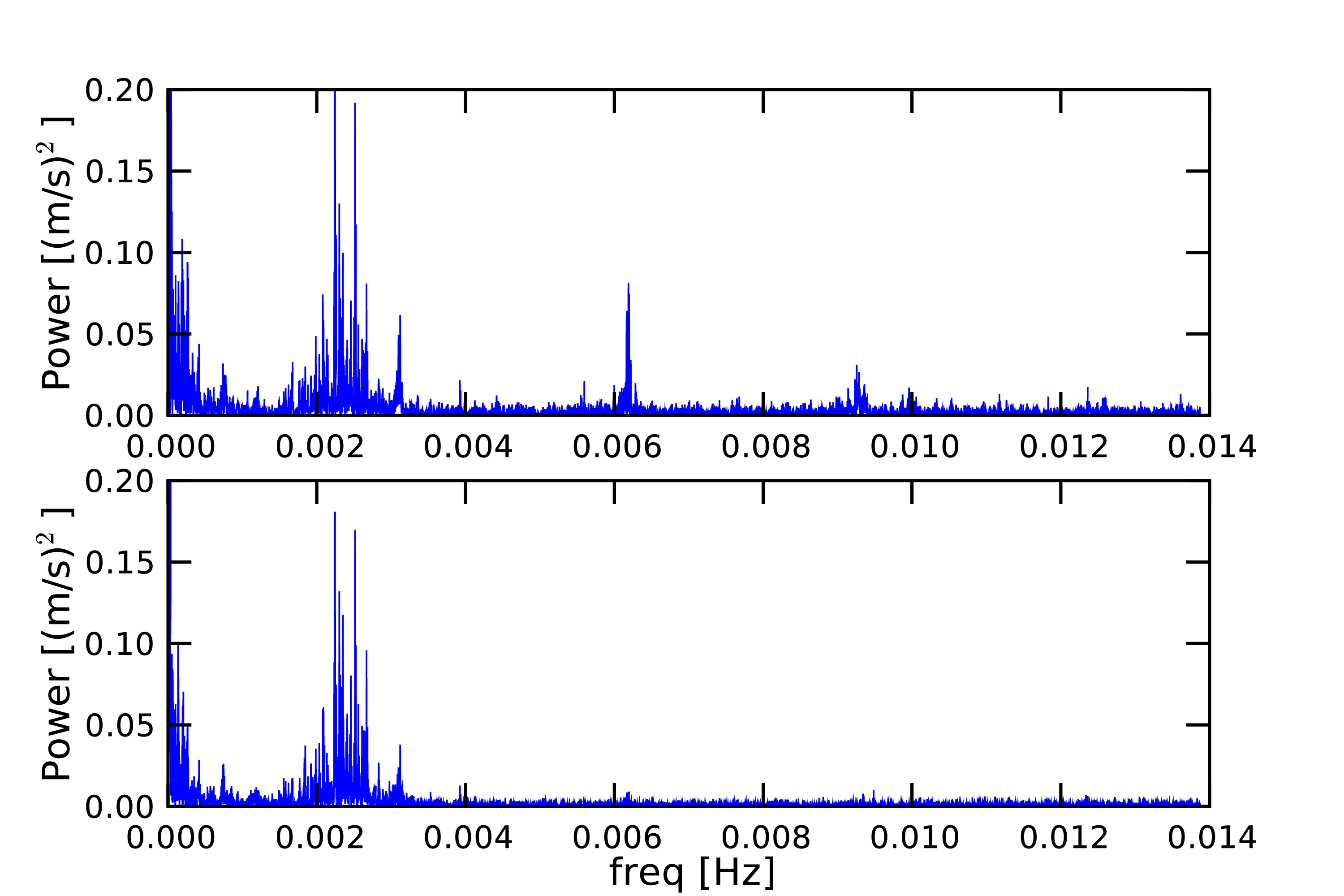}}
\caption{Cleaning of the guiding noise for $\alpha$\,Cen\,A. \emph{Top panel:}Power spectrum of the RV measurements, using the raw errors to calculate the statistical weights. We can see clearly the peaks at 3.1, 6.2 and 9.3 mHz due to the guiding noise.  \emph{Bottom panel:}Same power spectrum but after the cleaning process. The guiding noise peaks are much fainter.}
\label{fig:0}
\end{center}
\end{figure}

In the raw VPSD of all the stars, except $\alpha$\,Cen\,B, we notice huge excess of power at 3.1, 6.2 and 9.3 mHz (see Fig. \ref{fig:0}, top panel). It as been shown that these anomalies are due to a problem in the guiding engine of the telescope, resulting in a poor guiding precision \citep[e.g][]{Teixeira-2009,Bedding-2007b,Bazot-2007,Carrier-2006b}. The data for $\alpha$\,Cen\,B, taken in 2009, does not present these peaks due to corrections in the guiding software.

\begin{figure*}[!]
\begin{center}
\includegraphics[width=8cm]{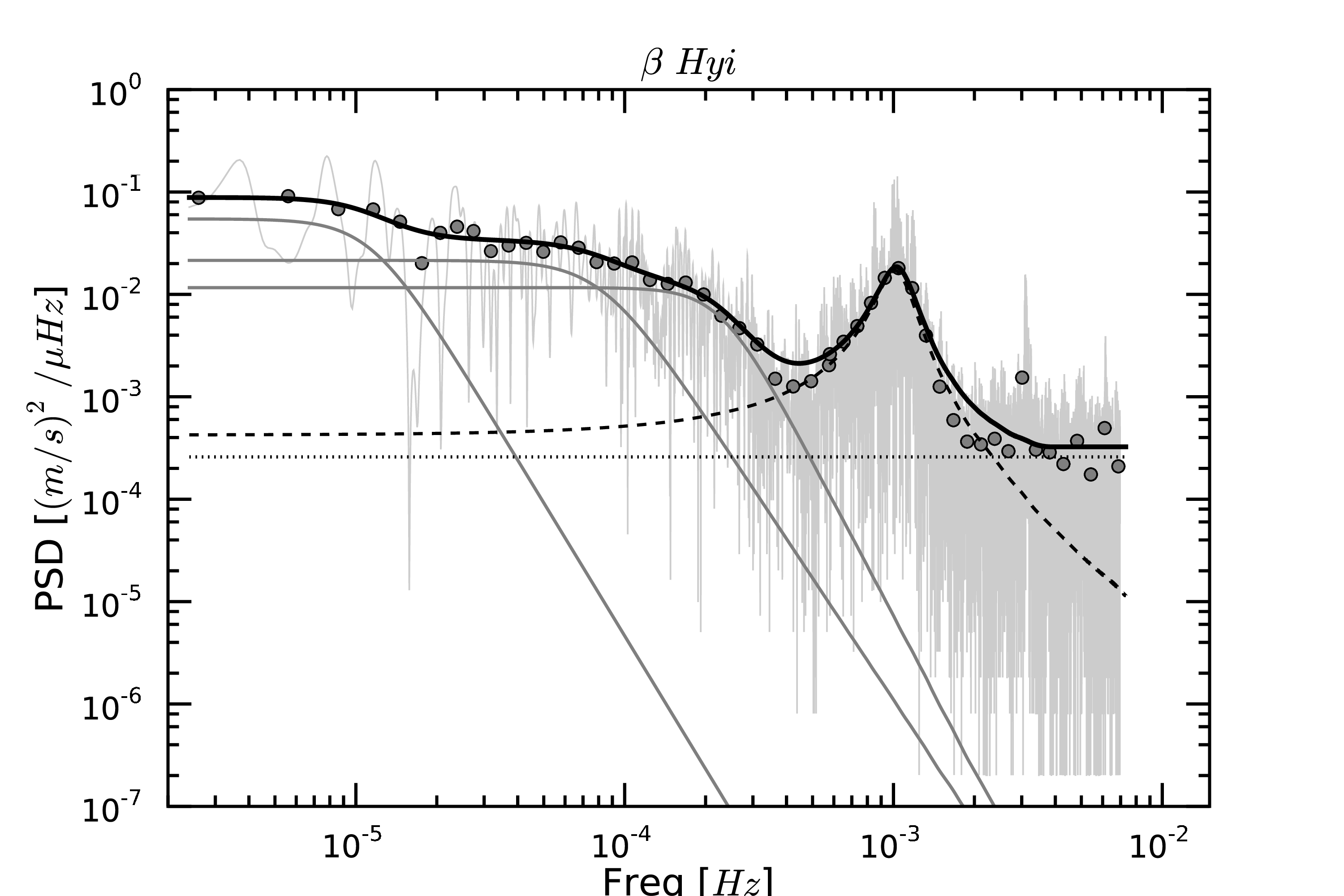}
\includegraphics[width=8cm]{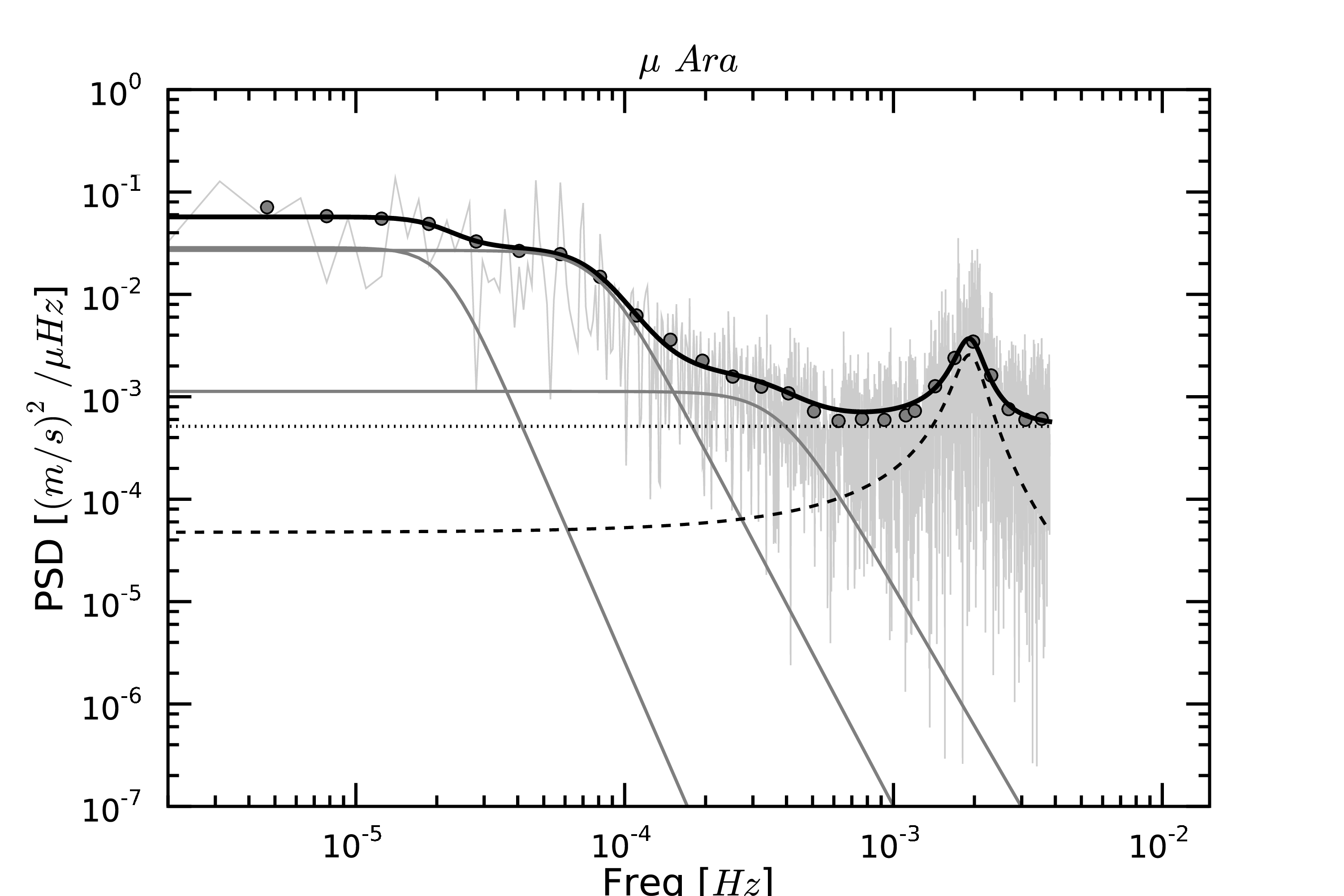}
\includegraphics[width=8cm]{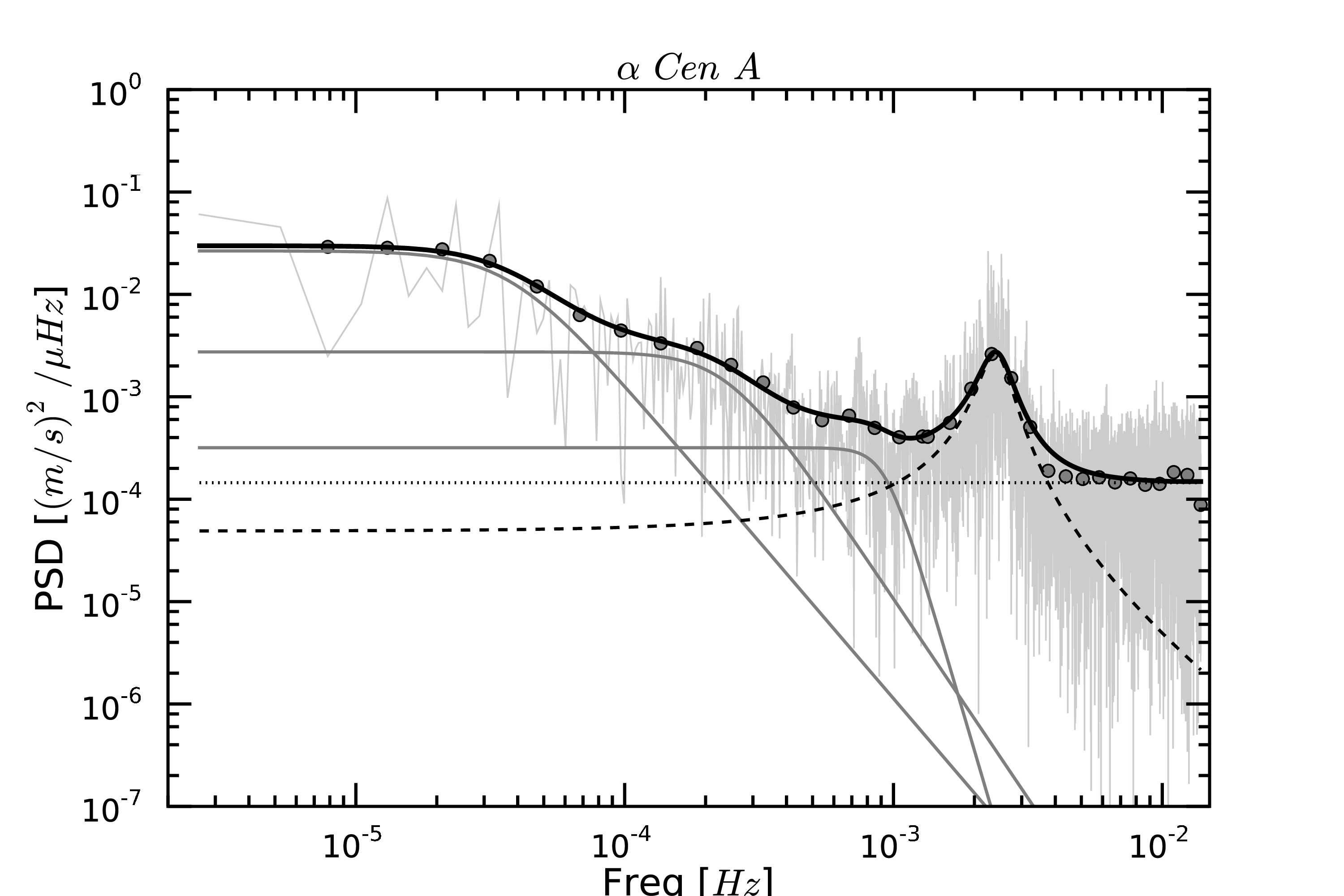}
\includegraphics[width=8cm]{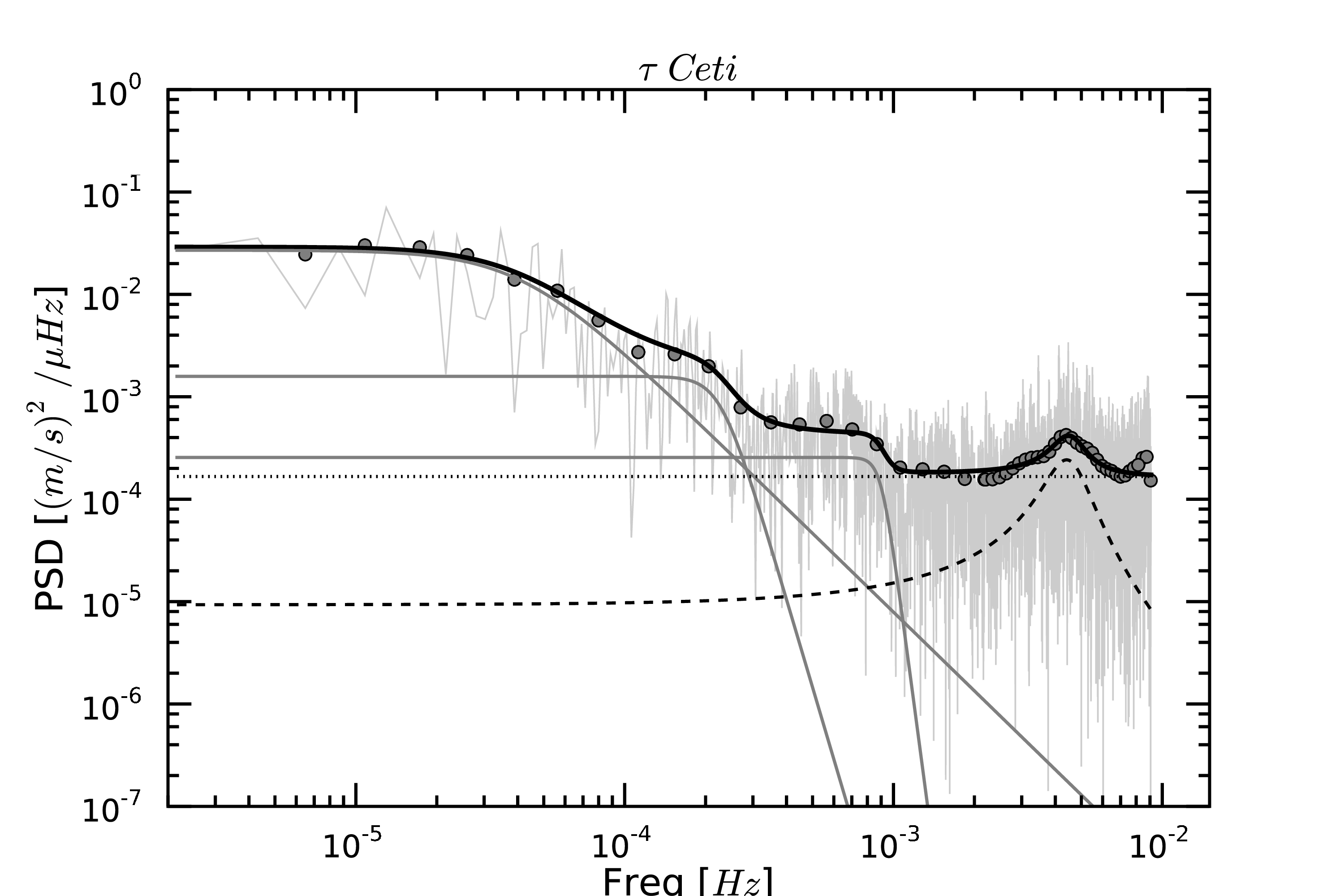}
\includegraphics[width=8cm]{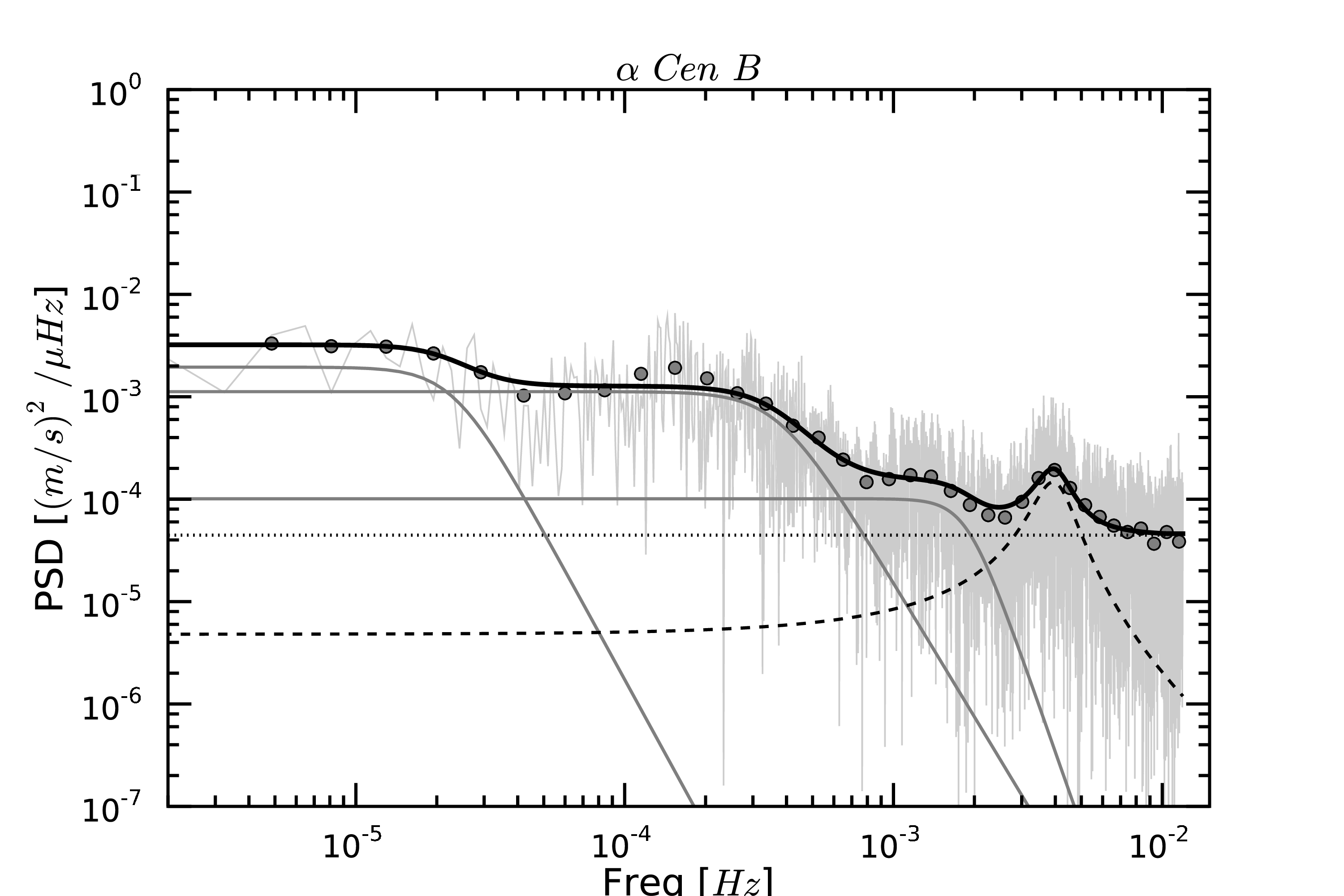}
\caption{VPSDs and  fits for the 5 stars of our samples. The grey dots corresponds to the binned spectrum using a boxcar algorithm with a changing width according to the frequency. The global fit including granulation phenomena, oscillation modes, photon noise and instrumental noise (black line) is obtained on the binned spectrum. Each contribution to the fit is shown. We can see the 3 types of granulation fitted in light grey lines (supergranulation, mesogranulation and granulation from left to right respectively). The Lorentzian fit to p-modes corresponds to the dashed line and the constant, fitting the photon and instrumental noises, can be see in dotted line.}
\label{fig:1}
\end{center}
\end{figure*}

To correct this noise effect in our measurements, we use the method developed for $\alpha$\,Cen\,A \citep[][]{Butler-2004b}, $\beta$\,Hyi \citep[][]{Bedding-2007b} and $\tau$\,Ceti \citep[][]{Teixeira-2009}. The idea is to down weight bad points produced by the poor guiding. First of all, we compute the raw power spectrum and remove the highest peaks corresponding to oscillations or granulation phenomena. We subtract the related sinusoid component from the RV data and we restart the process of calculating the periodogram and selecting the highest peak until all the power coming from real signals have been killed\footnote{For $\mu$\,Ara, $\alpha$\,Cen\,A and $\alpha$\,Cen\,B, all the peaks below 2.5, 3.5 and 6 mHz are removed, respectively. For $\tau$\,Ceti, following \citet{Teixeira-2009}, we kill all the frequencies below 0.8 mHz and between 3.1 and 5.5 mHz.}. The residual velocities obtained should only be composed of noise. If this noise is pure photon noise, the residual velocities should respect a gaussian distribution. For each night, we select all the point deviating from the mean scatter by more than 3$\sigma$ and we multiply their statistical weight by the probability of the point to belong to a gaussian distribution. Finally, we scale all the statistical weights, night by night, to reflect the noise measured at high frequencies. Figure \ref{fig:0} shows, for $\alpha$\,Cen\,A, the improvement brought by the cleaning process. After down weighting bad points, the noise level at high frequencies is reduced by 25 to 30\,\% for all stars, except $\alpha$\,Cen\,B that do not present this guiding noise effect in the raw data. In the case of $\beta$\,Hyi, the power spectrum has been kindly provided by Tim Bedding and without having access to the raw RVs, we could not correct for this guiding noise effect (the peaks at 3.1 and 6.2 mHz are visible on the $\beta$\,Hyi power spectrum in Fig. \ref{fig:1}). The corrected VPSD will be used for the following of the study.

For all the stars in our sample, we observe a raise of the VPSD towards low frequencies (see Fig. \ref{fig:1}) due to the nature of granulation phenomena. The decrease with frenquency follows until reaching the Lorentzian p-modes bump. Finally, after the signature of oscillation modes, we can see that the amplitude of the VPSD stays at the same level due to photon and instrumental noise.

\subsection{Fitting the velocity power spectrum density}\label{subsect:2.1}

Once the VPSD is calculated, we need to fit it to obtain a spectrum that is independent of the original data sampling. This will also allow us to separate individually each type of noise, which will give us an idea about the RV contribution of each of them.

The part of the VPSD corresponding to granulation phenomena can be explained by a model in which each source of convective motions is described by an empirical law initially proposed by \citet[][]{Harvey-1984} and reviewed by \citet[][]{Andersen-1994,Palle-1995}. This model corresponds to an exponentially decaying function with frequency:
\begin{equation}  \label{eq:1}
P(\nu)=\frac{A}{1+(B\nu)^C},
\end{equation}
where $P(\nu)$ is the VPSD, A the power density of the corresponding convection motion, B its characteristic time scale and C the slope of the power law. The power remains approximatively constant on time scales larger than B and drops off for shorter time scales.
The total VPSD due to granulation phenomena is the sum of the VPSD for each phenomenon (G = granulation, MG = mesogranulation and SG = supergranulaion):
\begin{equation} \label{eq:2}
P_{Tot}(\nu)=P_{G}(\nu)+P_{MG}(\nu)+P_{SG}(\nu).
\end{equation}

The frequency of the oscillation modes (p-modes) can analytically be derived using a perturbation theory and the asymptotic theory \citep[][]{Tassoul-1980}, leading to:
\begin{equation} \label{eq:3}
\nu_{nl}\sim \Delta\nu \left( n+\frac{l}{2}+\alpha \right),
\end{equation}
where $\Delta\nu$ is the large separation, $n$ the radial order and $l$ the angular degree of the mode. We are interested in the total power due to p-modes and not in each mode separately. We thus use the technique proposed by \citet{Kjeldsen-2005} \citep[see also][]{Arentoft-2008} to produce a single hump of excess power that is insensitive to the fact that the oscillation spectrum has discrete peaks. This is done by convolving the observed VPSD with a Gaussian having a full width at half maximum\footnote{The value of $\Delta\nu$ for each star of our sample is taken from the literature:57.2 $\mu Hz$ for $\beta$\,Hyi \citep{Bedding-2007b} , 106 $\mu Hz$ for $\alpha$\,Cen\,A \citep[][]{Kjeldsen-2005}, 90 $\mu Hz$ for $\mu$\,Ara \citep[][]{Bazot-2005}, 169 $\mu Hz$ for $\tau$\,Ceti \citep[][]{Teixeira-2009} and 162 $\mu Hz$ for $\alpha$\,Cen\,B \citep[][]{Kjeldsen-2005}} of $4\Delta\nu$. Finally, we fit the convolved curve with a Lorentzian function \citep[e.g.][]{Lefebvre-2008}:
%
%

\begin{equation} \label{eq:3.1}
P(\nu) = A_L\frac{\Gamma^2}{(\nu-\nu_0)^2+\Gamma^2},
\end{equation}
where $A_L$ correspond to the amplitude of the Lorentzian, $\Gamma$ to the full width at half maximum (FWHM) and $\nu_0$ to the mean of the Lorentzian.

At very high frequency, after the p-modes bump, we can see the contribution of photon and instrumental noises. Supposing that these noises are gaussian, which is at least the case for photon noise, their contribution correspond to a constant power ranging over all the spectrum.

To summarize, the entire VPSD can be adjusted by the following function with 13 free parameters:
\begin{equation} \label{eq:3.1}
P(\nu) = \sum_{i=1}^3{\frac{A_i}{1+(B_i\nu)^{C_i}}}+A_L\frac{\Gamma^2}{(\nu-\nu_0)^2+\Gamma^2}+constant,
\end{equation}
where $i$ correspond to each type of granulation phenomena. In order to perform the VPSD adjustment with 13 free parameters, we bin the spectrum, using a boxcar algorithm\footnote{The width of the box changes according to the frequency. The goal is to obtain a binned spectrum which has nearly the same spacing between the points in a logarithmic scale, which makes fit convergence faster.}, and use an iterative approach. We first adjust the Lorentzian and the constant on the oscillation modes, and then, fit the Harvey functions on the low frequency part of the spectrum, taking into account the constant derived before. Finally, we derive a global simultaneous fit of the model including the Harvey functions, the Lorentzian and the constant using as initial conditions the results of the individual fits. The VPSD and final adjustment for each star can be seen in Fig. \ref{fig:1} and the different parameters, in Table \ref{tab:1.1}.
\begin{table*}[!t] 
\begin{center}
\caption{Parameters for the spectrum fit (SG = supergranulation, MG = mesogranulation, G = granulation). A$_{SG}$, A$_{MG}$, A$_{G}$, A$_L$ and the constant are expressed in power density, $\frac{(m/s)^2}{\mu Hz}$.}  \label{tab:1.1}
\begin{tabular}{cccccccccccccc}
\hline
\hline Star & A$_{SG}$ & B$_{SG}$ [h] & C$_{SG}$ & A$_{MG}$ & B$_{MG}$ [h] & C$_{MG}$ & A$_{G}$ & B$_{G} [min] $ & C$_{G}$ &  $A_L$ & $\Gamma$ [mHz] & $\nu_0$ [mHz] & constant \\
\hline
$\beta$\,Hyi & 0.055 & 24.3 & 4.3 & 0.021 & 3.4 & 4.0 & 11.6\,10$^{-3}$ & 72.8 & 5.0 & 18.5\,10$^{-3}$ & 0.17 & 1.0 & 2.6\,10$^{-4}$ \\
$\mu$\,Ara & 0.029 & 13.0 & 6.0 & 0.027 & 3.4 & 5.0 & 1.1\,10$^{-3}$ & 43.8 & 4.5 & 3.2\,10$^{-3}$ & 0.26 & 1.9 & 5.1\,10$^{-4}$ \\
$\alpha$\,Cen\,A & 0.027 & 7.4 & 3.1 & 0.003 & 1.2 & 3.9 & 0.3\,10$^{-3}$ & 17.9 & 8.9 & 2.6\,10$^{-3}$ & 0.36 & 2.4 & 1.4\,10$^{-4}$ \\
$\tau$\,Ceti & 0.027 & 6.7 & 2.6 & 0.002 & 1.2 & 8.9 & 0.3\,10$^{-3}$ & 18.5 & 19.8 & 0.3\,10$^{-3}$ & 0.75 & 4.5 & 1.7\,10$^{-4}$ \\
$\alpha$\,Cen\,B & 0.002 & 12.0 & 4.8 & 0.001 & 0.7 & 4.4 & 0.1\,10$^{-3}$ & 8.9 & 7.5 & 0.2\,10$^{-3}$ & 0.68 & 3.9 & 0.5\,10$^{-4}$ \\
\hline
\end{tabular} 
\end{center}
\end{table*}

For the case of $\alpha$\,Cen\,B, the observation data we have does not correspond to continuous follow up of the star. To check if it does not affect high frequency noises, we compare the high frequency spectrum of our data with one night of asteroseismology done on $\alpha$\,Cen\,B in 2004 (HARPS ESO archive). We find the same p-modes envelope in both case, which proves that the data we have for $\alpha$\,Cen\,B are as good as the one gathered for the 4 other stars (see Fig. \ref{fig:1}).

Concerning oscillation modes, all the stars of our sample have been previously studied in details. We thus can compare the mean frequency of the p-modes bump fitted in this paper, with the one found in previous studies \citep[][and reference therein]{Arentoft-2008,Kjeldsen-2008}. According to Fig. 11 in \citet{Arentoft-2008}, the mean frequency of the p-modes bump are the following: 1\,mHz for $\beta$\,Hyi, 2\,mHz for $\mu$\,Ara, 2.5\,mHz for $\alpha$\,Cen\,A, 4.5\,mHz for $\tau$\,Ceti and 4\,mHz for  $\alpha$\,Cen\,B. These frequencies can be compared with our results, see $\nu_0$ values in Table \ref{tab:1.1}, and we find a very good agreement.

To compare the fitted parameters of granulation phenomena with the literature values is more complex. To our knowledge, only the Sun has been previously studied in detail \citep[][]{Lefebvre-2008,Palle-1995}. In addition, the normalization of the algorithms used by these authors to calculate the VPSD is different from the one adopted here. This implies the derivation of different amplitudes for the considered granulation phenomena. Some work could be done to normalize the values, but this is out of the scope of the present paper. We can nevertheless compare the different time scales obtained. In \citet{Lefebvre-2008} and \citet{Palle-1995}, we have access the the characteristic time $\tau$, which is by definition equals to $B/(2\pi)$. Changing the characteristic times from $\tau$ to $B$, \citet{Lefebvre-2008} obtain $B_G$ = 22\,min whereas \citet{Palle-1995} get $B_G$ = 39\,min. In our case, for $\alpha$\,Cen\,A (G2V) and $\mu$\,Ara (G3IV-V) which are close to the Sun in terms of spectral class, we find $B_G$ = 18\,min and 44\,min respectively, which is in good agreement. For the case of meso- and supergranulation, we find some discrepancies in the literature. In \citet{Palle-1995}, $B_{MG} \sim 17\,h$ and $B_{SG} \sim175\,h $. According to more precise and recent results, we can extract roughly from Fig. 2 in \citet[][]{Lefebvre-2008} that $B_{SG} \sim 2.10^{-5}\,Hz \sim 14\,h$. Notice that our results for $\alpha$\,Cen\,A (G2V) and $\mu$\,Ara (G3IV-V) are close to this second value.


\subsection{Synthetic radial velocity measurements}\label{subsect:2.3}

Once the VPSD fit is obtained, we have to calculate the RV for each date of an observational calendar. Since the periodogram technique is equivalent to a weighted sine wave fitting \citep[][]{Scargle-1982,Zechmeister-2009}, to pass from the spectrum to RVs, we have to calculate for the desired epochs $t_i$: 
\begin{equation} \label{eq:4}
RV(t_i)=\sum_{\nu}{\sqrt{\textrm{VPSD}(\nu)}(\sin(2\pi\nu t_i+phase(\nu)))},
\end{equation}
where the frequency $\nu$ goes from 1/T to the Nyquist frequency with a step of 1/T. Supposing that all the noises due to stellar perturbations are independent, we choose a uniformly random phase\footnote{Using the phase calculated by the periodogram is not what we want because it will reproduce the real data, setting the RVs at 0 during the observation gaps.} between 0 and $2\pi$.

To check if the computation of synthetic RV measurement over long period does not introduce any noise, absent from the asteroseismology measurements, we compute the rms of synthetic RV measurements for different time spans. To be consistent with the observational strategies we will use afterwards, we calculate the rms for 1 measurement per night of 15 minutes for 10, 50, 100 and 1000 consecutive nights. For each star, each measurement of 15 minutes is divided in several exposures according to the exposure time of the asteroseismology run (see Table \ref{tab:1}). The difference between the synthetic RV rms for each time span and the rms of the real RV is less than 10\,\% for all the stars. This corresponds to less than 25\,cm\,s$^{-1}$ which is significantly below the long term precision level of HARPS (60-80 cm\,s$^{-1}$, HARPS team, private communication). This proves that computation of synthetic measurements over long period does not introduce any bias. Therefore, such measurements contain only short term perturbations (oscillations, granulation phenomena, photon and instrumental noise and a little bit of activity) even if the total time span is much longer.

Table \ref{tab:2} gives the rms for the real RVs, $rms_{data}$ and for the synthetic RVs for 100 consecutive nights, $rms_{syn,\,tot}$. We also calculate the rms expected for each noise contribution as well as the high frequency noise derived using the power spectrum. This high frequency noise corresponds to the median level of the power spectrum for frequencies higher than p-modes. We notice a good agreement between the rms calculated using the fitted constant and using the power at high frequency, which guaranty a correct estimation of the photon and instrumental noise.
\begin{table*}[!t] 
\begin{center}
\caption{Comparison of the dispersion of real RVs, $rms_{data}$, with the synthetic data dispersions for all type of noise, $rms_{syn,\,tot}$, and for each type of noise individually (SG = supergranulation, MG = mesogranulation, G = granulation, const = instrumental and photon noise). The dispersion of synthetic measurement is calculated on 100 consecutive days, with 1 measurements per night of 15 minutes. $rms_{high\,frequency}$ corresponds to the noise calculated in the power spectrum at high frequency, after p-modes.}  \label{tab:2}
\begin{tabular}{ccccccccc}
\hline
\hline Star & $rms_{data}$ & rms$_{syn,\,tot}$ & rms$_{syn,\,SG}$ & rms$_{syn,\,MG}$ & rms$_{syn,\,G}$ & rms$_{syn,\,p-modes}$ & rms$_{syn,\,const}$ & rms$_{high\,frequency}$\\
 & [m\,s$^{-1}$] & [m\,s$^{-1}$] & [m\,s$^{-1}$] & [m\,s$^{-1}$] & [m\,s$^{-1}$] & [m\,s$^{-1}$] & [m\,s$^{-1}$] & [m\,s$^{-1}$]\\
\hline
$\beta$\,Hyi & 2.60 & 2.88 & 0.55 & 1.05 & 1.26 & 2.23 & 0.94 & -\\
$\mu$\,Ara & 1.97 & 1.90 & 0.56 & 1.05 & 0.46 & 1.11 & 1.00 & 1.08\\
$\alpha$\,Cen\ A & 2.03 & 1.89 & 0.77 & 0.62 & 0.40 & 1.20 & 0.99 & 1.1\\
$\tau$\,Ceti & 1.47 & 1.42 & 0.85 & 0.43 & 0.34 & 0.52 & 0.87 & 0.96\\
$\alpha$\,Cen\,B  & 0.9 & 0.88 & 0.15 & 0.48 & 0.32 & 0.39 & 0.52 & 0.52\\
\hline
\end{tabular}
\end{center}
\end{table*}

\begin{figure}
\resizebox{\hsize}{!}{\includegraphics{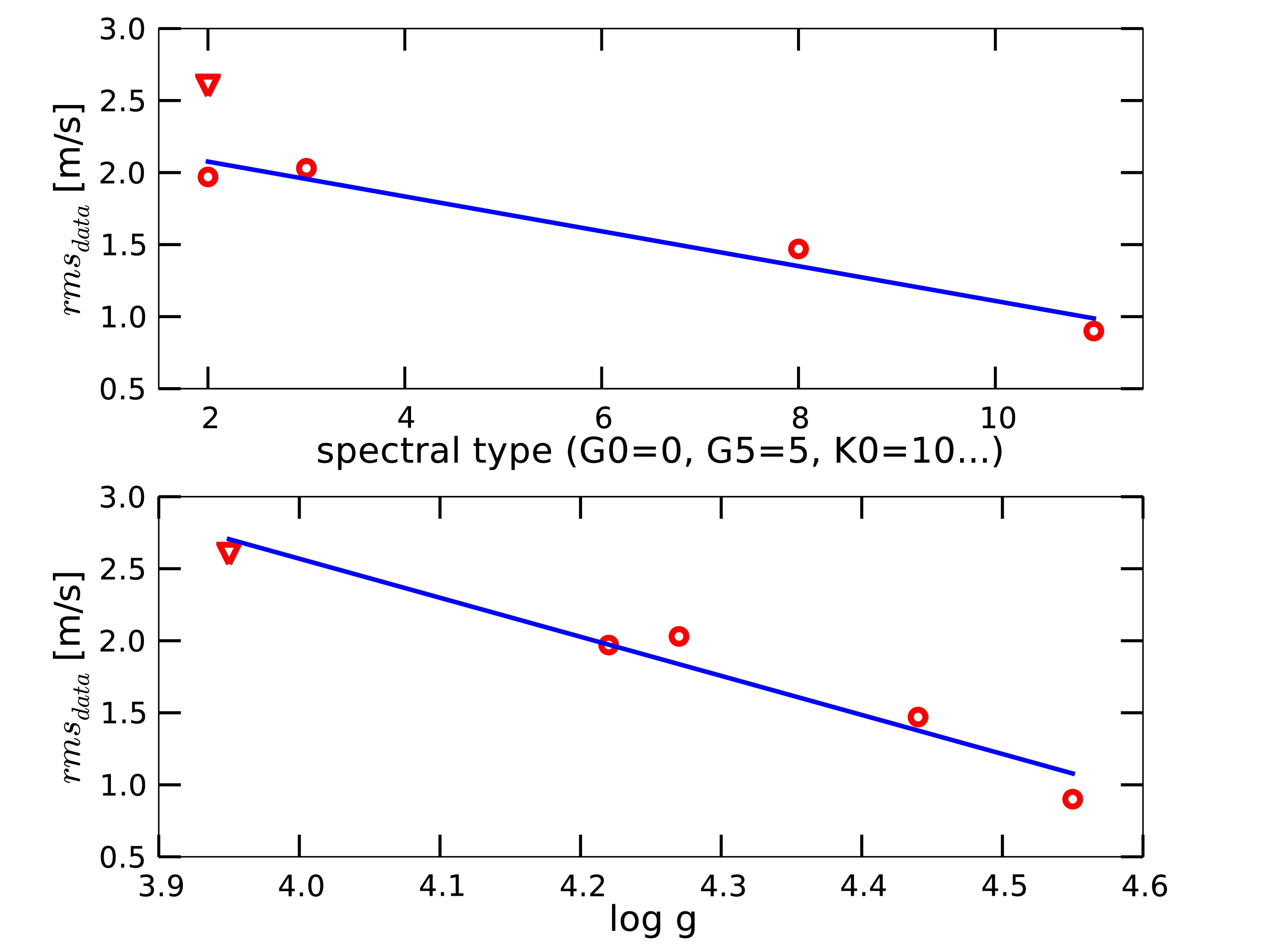}}
\caption{\emph{Top panel:} RV variation, $rms_{data}$, in function of the spectral type. The fit is done on dwarf stars (circles) without taking into account the sub-giant (triangle). \emph{Down panel:} RV variation in function of the surface gravity, log g. The linear fit is done on all the sample.}
\label{fig:4.2}
\end{figure}

As already pointed out in \citet{Mayor-2003b}, we observe for dwarf stars a trend between the level of RV variation and spectral type. If we look at the main sequence stars in our sample ($\alpha$\,Cen\,A (G2V), $\mu$\,Ara (G3IV-V), $\tau$\,Ceti (G8V) and $\alpha$\,Cen\,B (K1V)), we notice a decrease of the level of RV variation from G to K dwarfs. Figure \ref{fig:4.2} (top panel) illustrates well this trend. A similar behaviour can also be pointed out when comparing the RV variation with surface gravity, log g. In this case, all the stars follow the trend, even $\beta$\,Hyi (G2IV) which is a sub giant. Thus, it seems that the RV variation is dependent on the spectral type and the evolutionary stage of the star. The expected results on averaging the stellar noise will then strongly depend on the characteristics of the star considered.

Since the fit is the sum of each noise contribution, see Eq. \ref{eq:3.1}, we can extract each noise independently from the VPSD and calculate the corresponding rms. Except for supergranulation, we can see in Table \ref{tab:2} that the noise level of granulation phenomena, as well as oscillations, decreases when we go towards late spectral type stars (the stars are ordered from early spectral type to late spectral type and from evolved star to non evolved). The calculation of the noise rms induced by supergranulation is very sensible to the parameters B and C as well as the minimum frequency of the spectrum, which explains why we do not see any trend between the noise level of supergranulation and the spectral type. The effect of noise amplitude in function of the spectral type and/or evolution is better seen in Table \ref{tab:1.1} by comparing the values for A$_{SG}$, A$_{MG}$, A$_{G}$ and A$_{L}$. In this case, we can see the dependency of supergranulation. \citet{Kjeldsen-1995}, \citet{Christensen-Dalsgaard-2004} and \citet{OToole-2008} already showed the dependency between amplitude of stellar noise and spectral type and evolution for oscillation modes. At our knowledge, it is the first time that this dependency is shown for granulation phenomena. In conclusion, early K dwarfs have a total noise level lower than G dwarfs or sub-giants, making them better candidates for the investigation of exoplanets using precise radial-velocity measurements.

All the stars in our sample are very bright, below magnitude 5.2 (see Table \ref{tab:1}), which induce a very small photon noise. For $\mu$\,Ara which is the fainter one, the photon noise estimated for a single exposure of the asteroseismology run is below 34 cm\,s$^{-1}$. Comparing this value with the one found for the photon and instrumental noise (see rms$_{syn,\,const}$ in Table \ref{tab:2}), we see that the precision is limited by the instrument. We can separate our sample in 2. On one side $\alpha$\,Cen\,B which as a very low photon and instrumental noise, and on the other side, the other stars. Measurements for $\alpha$\,Cen\,B are very recent and a lot of work as been done to reduce the effect of guiding noise, which explains the rms difference between the 2 samples. In the sample with high rms, all the stars seem to have a similar instrumental noise, around 0.9 to 1 m\,s$^{-1}$. Several factors can be responsible for the small rms difference. For example, the quality of the sky plays an important role in the instrumental noise, mainly guiding noise. If we look at a star with a very good seeing, smaller than the spectrograph fiber diameter (1 arcesec), the point spread function of the star will not cover the entire fibre. This will introduce a noise when the star will change of position on the fiber because of imperfect scrambling. This noise can be average for long exposure times, but not for high observational frequency as requested by asteroseismology.

\subsection{Correcting instrumental noise for long period}\label{subsect:2.4}

As shown in last paragraph, our RV measurements seem instrumental noise limited. This noise is coming mainly from the telescope guiding system which is not optimized to observe stars using very short exposure times. For planet surveys, we normally use exposure times of 15 minutes which allow a good averaging of the guiding noise and should give smaller instrumental noise. Moreover, estimating the instrumental noise with asteroseismology measurements gives a good approximation of the noise on short time scales, but not on the long time scales required for planet surveys. For long time scales and 15 minutes exposure time per measurement, the instrumental noise on HARPS is estimated to be between 0.6 and 0.8 m\,s$^{-1}$ (HARPS team, private communication). To take into account long term instrumental effects and remove short term perturbations due to guiding, keeping the same contribution from oscillations and granulation phenomena, we readjusted the level of the fitted constant to 0.8\,m\,s$^{-1}$ for all the stars and recalculated the synthetic RVs. This value of 0.8\,m\,s$^{-1}$ has been taken to be conservative.

\section{Considered observing strategies}

Using the synthetic RVs for each of the considered stars, the goal is now to estimate, for different observational strategies, the variation level of binned RVs induced by stellar noise.

\subsection{Present HARPS-GTO observational strategy}\label{sect:3}

In order to be useful in a practical way, simulations have to take into account a realistic access to telescope time. A suitable approach is provided by the scheduling of the high-precision HARPS-GTO program. A star like HD\,69830 \citep[see][Lovis et al. 2010 in preparation for additional points]{Lovis-2006} as been followed using an ideal calendar of 1 measurement per night on 10 consecutive nights per month and over more than 4 years. In the present study, we will use this ideal calendar on 4 years and remove 4 month per year, since most of the stars disappear from the sky during this period. In addition, we suppress randomly 20\,\% of nights, to take into account bad weather or technical problems. Such a calendar represent thus 256 nights of measurements over a time span of 4 years.


For each star in our sample, we calculate the RV expected for each date of the calendar using an exposure time of 15 minutes, corresponding to the present HARPS-GTO observational strategy. This strategy will be referred hereafter as the 1N strategy. Then we make bins of 2, 5 and 10 consecutive nights and calculate the rms of the binned RVs, $rms_{RVb}$, in function of the binning in days.

\begin{figure*}
\centering
\includegraphics[width=9cm]{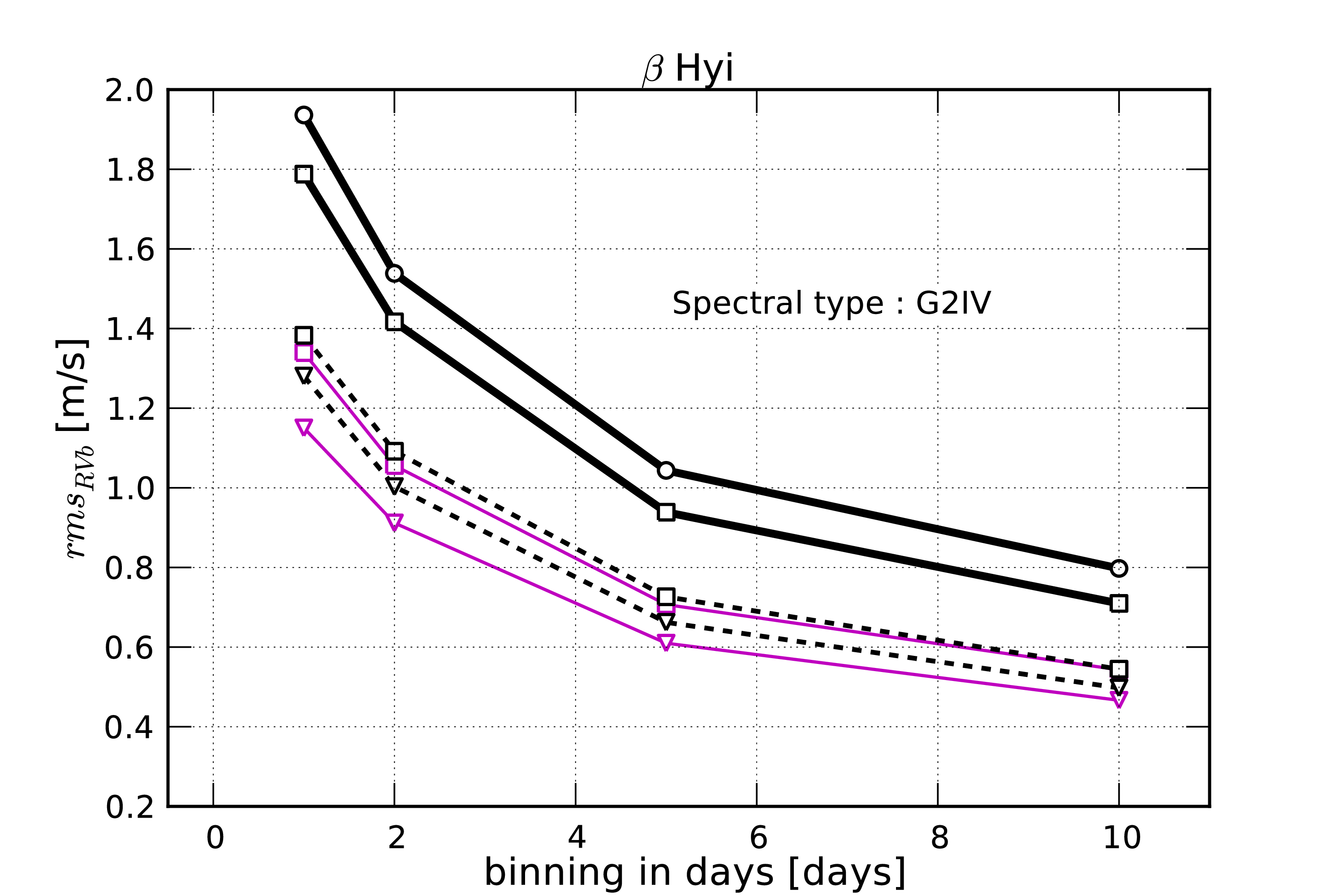}
\includegraphics[width=9cm]{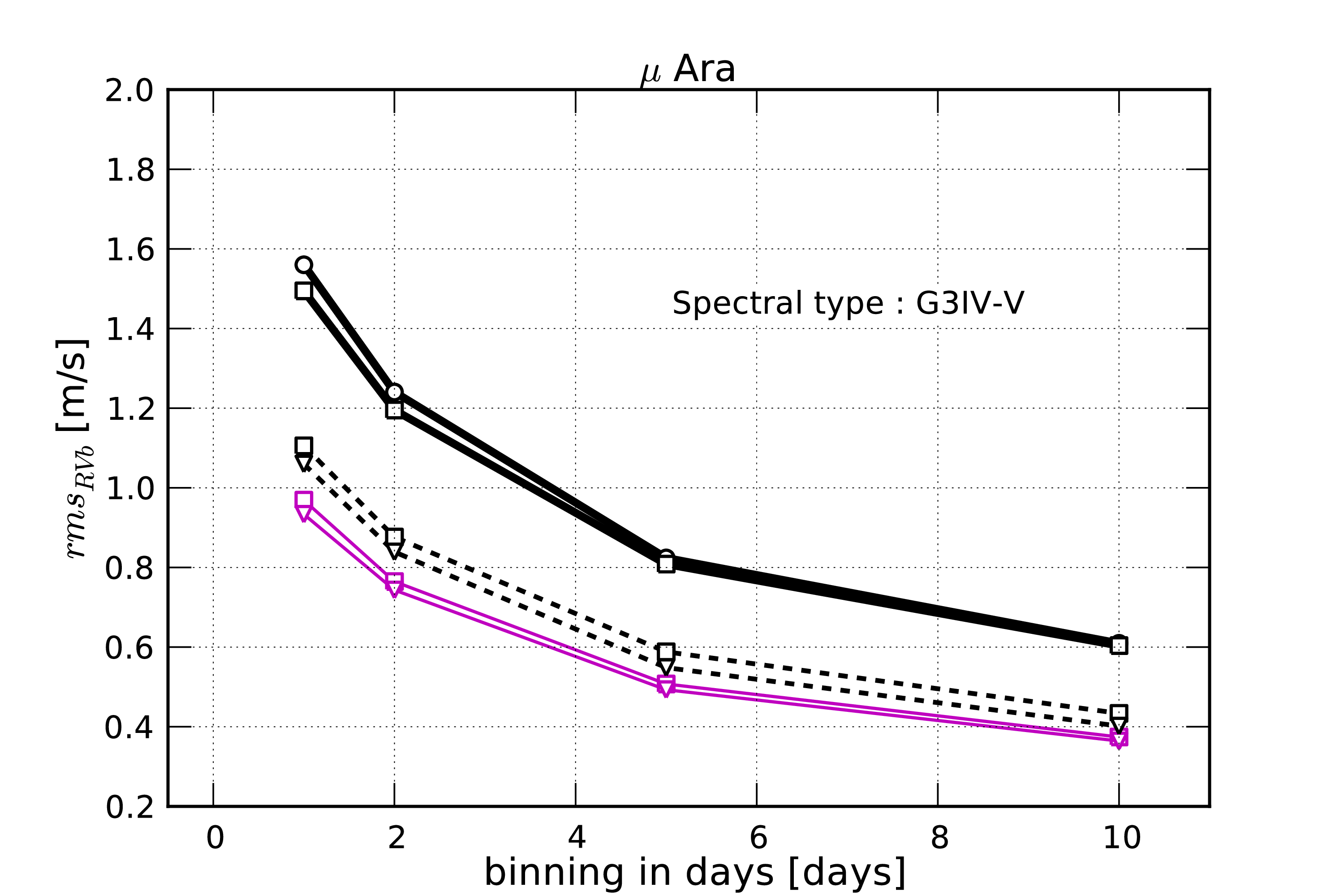}
\includegraphics[width=9cm]{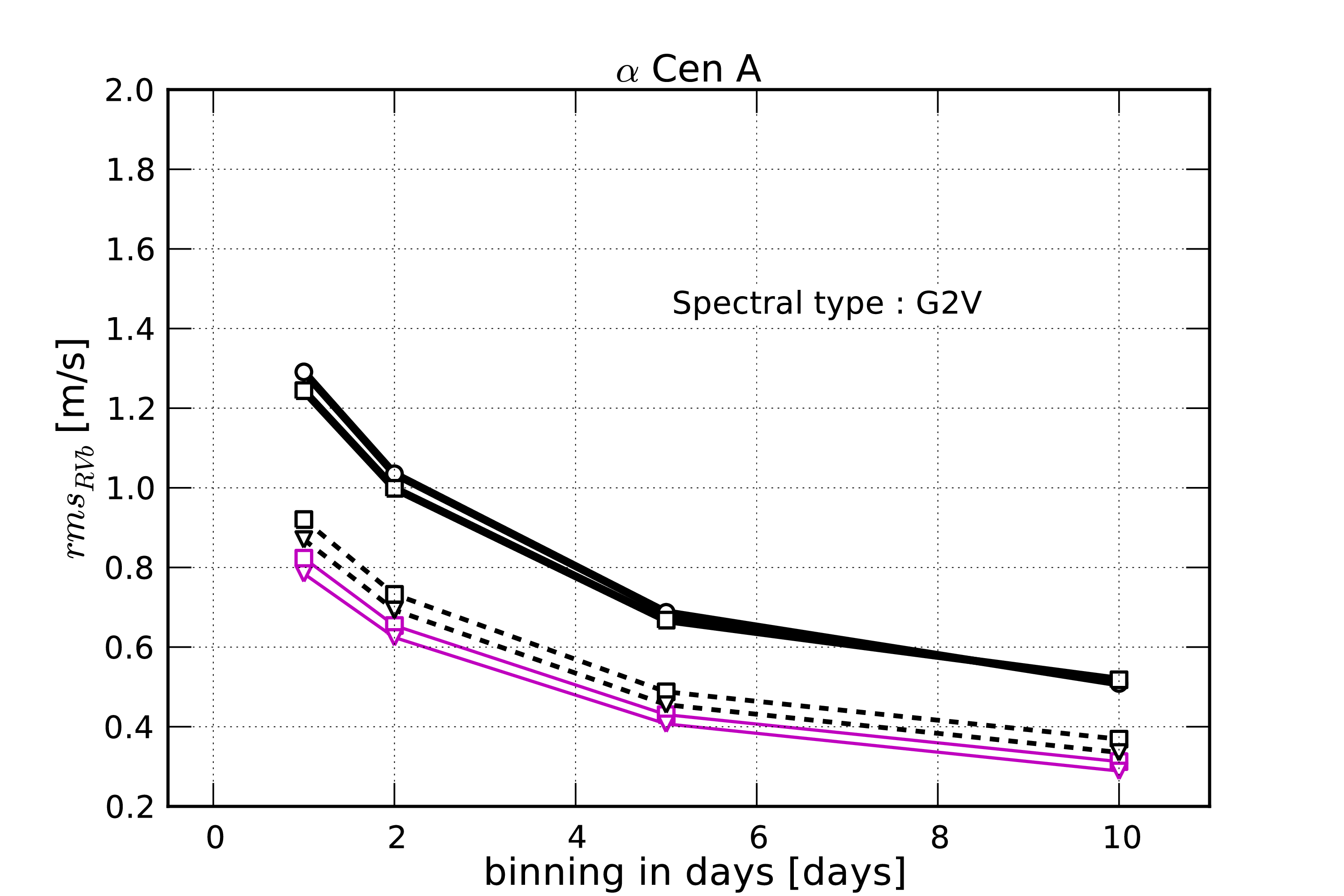}
\includegraphics[width=9cm]{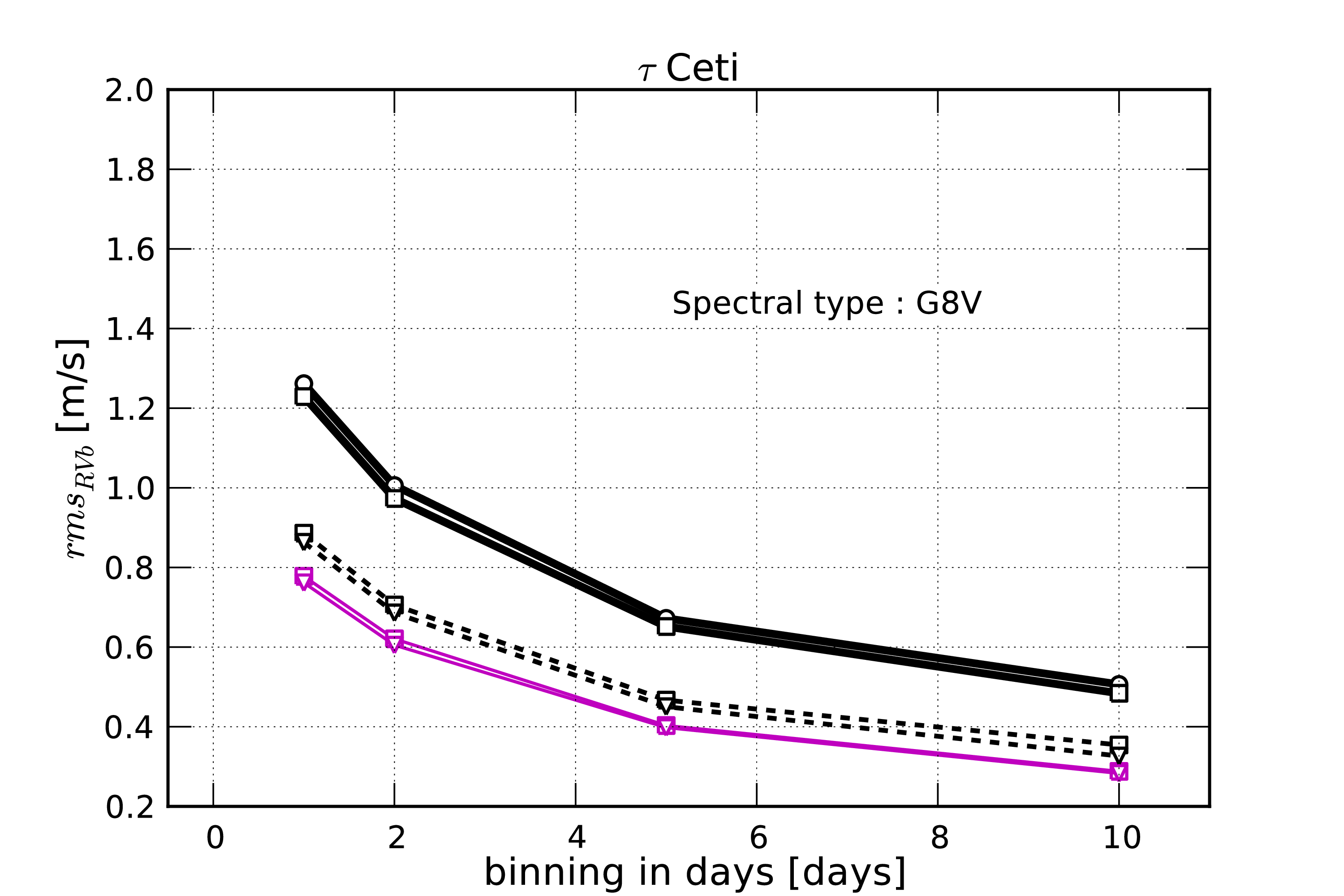}
\includegraphics[width=9cm]{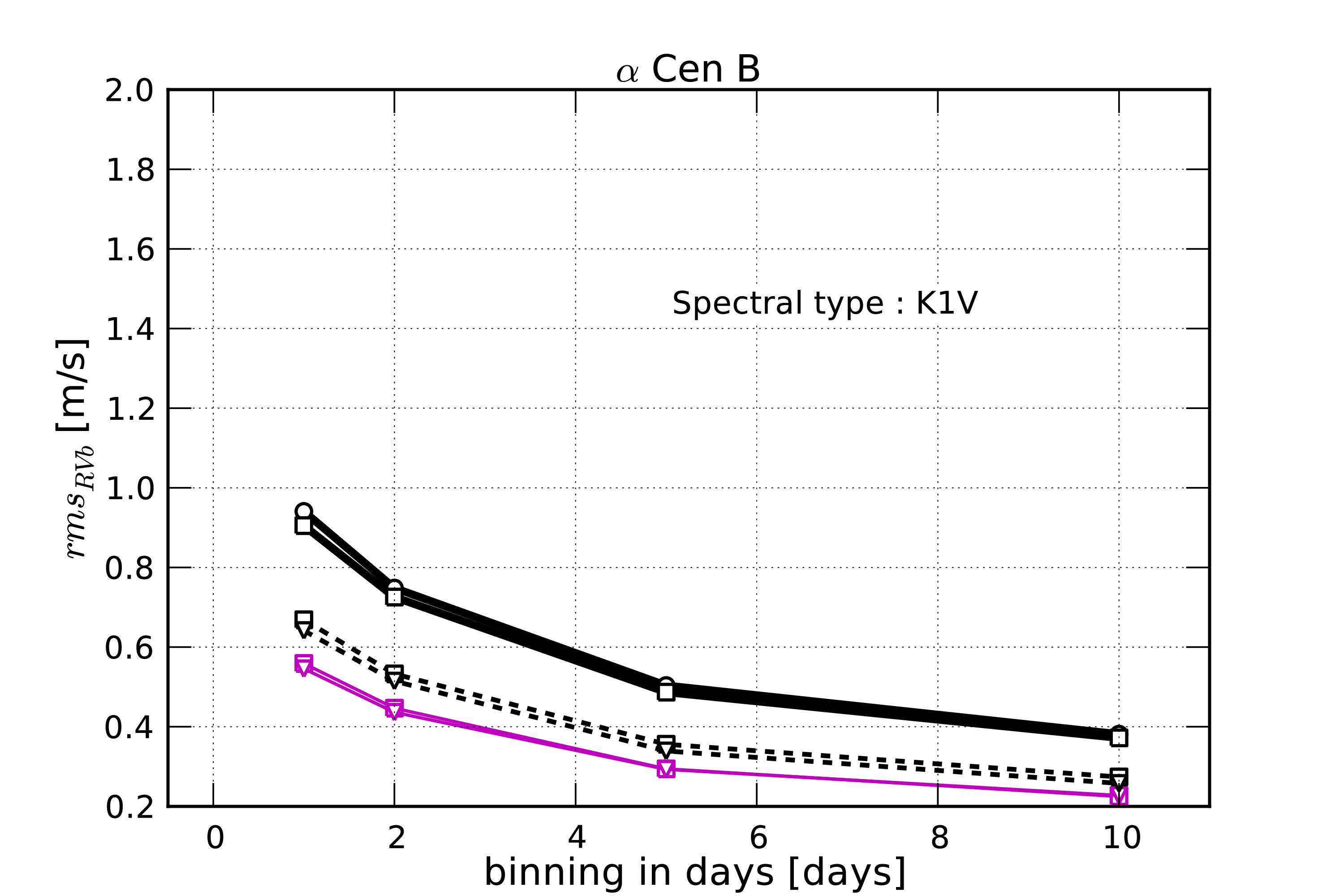}
\caption{Rms of the binned RVs, $rms_{RVb}$, as a function of the binning in days. The different considered strategies are 1 measurement per night (the 2 heavy continuous lines), 2 measurements per night with a spacing of 5 hours (the 2 dashed lines) and 3 measurements per night with a spacing of 2 hours (the 2 thin continuous lines). Circles represent a total observing time of 15 minutes per night (1$\times$15 minutes), squares represent a total observing time of 30 minutes per night (1$\times$30 minutes, 2$\times$15 minutes, and 3$\times$10 minutes) and triangles represent a total observing time of 1 hour per night (2$\times$30 minutes and 3$\times$20 minutes).}
\label{fig:4}
\end{figure*}

The result for each star is illustrated by the heavy solid black lines with circle markers on Fig.\,\ref{fig:4}. For example, in the case of  $\alpha$\,Cen\,B, we can see that with a binning in days of 10, we can reduce the RV variation, mainly due to stellar noise, down to the level of 0.38\,m\,s$^{-1}$, to be compared with 0.94\,m\,s$^{-1}$ in the case without binning. 
%

\subsection{Setting new observational strategies}\label{sect:4}

We have presented in the previous section the result expected when using the current HARPS observational strategy for the high-precision subprogram (1 measurement per night of 15 minutes). The goal of our work is to determine if we can reduce even more the noise contribution by sampling better the time scales of the various intrinsic stellar noise sources. The following additional strategies have been tested:  3 measurements per night with 2 hours of spacing and 2 measurements per night 5 hours apart. For the first strategy, we tested for each exposure 10 and 20 minutes and for the second one, 15 and 30 minutes. In addition a strategy with 1 measurement per night of 30 minutes has been tested. For long-period planet investigation, we moreover need to follow the star as long as possible along the year in order to obtain a good coverage in phase. Also, because of airmass limitations, we consider that stars can only be observed at most 5 hours per night under good conditions. This explains the choice of the selected approaches. Table\,\ref{tab:3} presents the different strategies used in the simulations.
\begin{table}[!t]
\begin{center}
\caption{Characteristics of the different observational strategies}  \label{tab:3}
\begin{tabular}{cccc}
\hline
\hline Measure & Number of & Exposure & Total time\\
spacing $[h]$ & meas./night & time $[min]$ & per night $[min]$\\
\hline
2 & 3 & 10 & 30\\
2 & 3 & 20 & 60\\
5 & 2 & 15 & 30\\
5 & 2 & 30 & 60\\
24 & 1 & 15 & 15\\
24 & 1 & 30 & 30\\
\hline
\end{tabular} 
\end{center}
\end{table}

The strategy with 15-minutes total exposure time per night corresponds to what is done in the high-precision HARPS-GTO subprogram. The corresponding results will be useful for comparison with results obtained with the new considered approaches. The different exposure times considered will give us a good feeling on how to average stellar oscillation effects (p-modes that have a typical time scale of a few minutes for solar type stars), whereas the multiple exposures during the night will tell us about possible dumping, after binning, of granulation phenomena effects at lower frequencies.

\section{Results for the selected observational strategies}\label{sect:5}

In this section, we focus on the improvement brought by the new strategies. First, we calculate the level of RV noise, $rms_{RVb}$, as a function of the binning in days. We use the same calendar as in the previous section, which will allow us to compare the new strategies with the present one used on HARPS.

\subsection{An efficient and affordable observational strategy}\label{subsect:5.1}

The evolution of the RV variation, $rms_{RVb}$, as a function of the binning in days is shown in Fig. \ref{fig:4}. In these plots, each type of line corresponds to the same number of measurement per night. For each of them, we have 2 curves, the lower one is obtained with an exposure time twice longer than the upper one. The difference between these 2 curves is small, so doubling the exposure time does not strongly improve the results, although it doubles the total measurement cost (neglecting overheads). Thus, it appears that the exposure time is not the only parameter that can average out stellar noise. The frequency of measurements plays also an important role.

Taking only the shortest exposure times, we are left with 3 strategies: 3 times 10-minute measurements per night with 2 hours of spacing between them (hereafter, 3N strategy), 2 times 15-minute measurements per night separated by 5 hours (hereafter, 2N strategy), and 1 measurement per night with 15 minutes exposure time (hereafter, 1N strategy). Comparing the 1 measurement-per-night of 30 minutes strategy with the 2N and 3N strategies allows to clearly see that with a similar observation time per night (30 minutes), splitting the measurements over the night in 2 or 3 blocks improve significantly the averaging of the stellar noise. The best among the considered strategies is the 3N strategy. This strategy gives values for $rms_{RVb}$ in average 30 to 40\,\% smaller than the present strategy used on HARPS for high precision.

As already pointed before, the RV noise due to stellar intrinsic perturbations is lower for K than for G dwarfs. Within a given spectral type, we can also compare this value between a main sequence and a more evolved star ($\alpha$\,Cen\,A (G2V) and $\mu$\,Ara (G3IV-V) or $\beta$\,Hyi (G2IV)). For more evolved stars, the RV variation seems higher, though we need to be careful since we just have one target, $\beta$\,Hyi ($\mu$\,Ara is between a dwarf and a subgiant).


\subsection{Strategy effects for the different stellar noise sources}\label{subsect:5.2}

The impact of exposure time and frequency of measurements on each kind of intrinsic stellar noise can be obtained by considering separately each noise contribution in the power spectrum. Since we fit each type of noise, we can extract independently the 3 types of granulation phenomena noises and the noise coming from oscillation modes. Then, using the technique to obtain synthetic RV measurements, we can create some data containing each type of noise, apply different strategies on them and thus see the impact of exposure time and frequency of measurements on each kind of stellar noise.

Without surprise, p-modes, causing oscillations of the solar surface over time scales smaller than 15 minutes can be adequately averaged by longer exposure times\footnote{We can take 1 measurement per night of 15 minutes or 3 measurements per night of 5 minutes and do some binning, the results are similar}. For slightly lower frequency noises, it is useful to increase the total exposure time as much as possible ($\sim$ 30 minutes). This will average the effect of granulation whose typical time scale is less than $\sim$25 minutes. For the mesogranulation, a further increase of the exposure time is poorly efficient whereas observing the star more often during the night leads to better results after binning. The 3N strategy is the one providing the best results for mesogranulation, since the 2 hours of spacing between the measurements over the night manages well to sample the perturbation time scale of the phenomenon. Finally, for supergranulation at even longer time scales, taking 2 measurements with a spacing of 5 hours gives the best results.


\section{Comparison with real observation}\label{sec:6}

Our simulation only takes into account oscillations, granulation phenomena, instrumental and photon noises. Lower frequency variations, due to active regions, are supposed to produce a higher level of noise, even for stars in their minimum of activity \citep[][]{Meunier-2010}. To check if our simulation, not considering active regions, corresponds to a realistic case, we have to compare the level of noise simulated with long term jitter observed on other stars.

Our simulation gives us the noise expected for a given observational strategy. In order to compare this simulated noise with real observations, we have to use an observational strategy similar to the one nowadays used on HARPS. We thus choose the 1N strategy. Using this strategy on $\alpha$\,Cen\,A and $\alpha$\,Cen\,B with no binning, the dispersion taking into account stellar, instrumental and photon noise is equal to 1.29 and 0.94\,m\,s$^{-1}$, respectively. These values can be compared to the rms of the residuals obtained for detected planetary systems. Table \ref{tab:0} gives the rms of the residuals for host star presenting a spectral type similar to $\alpha$\,Cen\,A (G2V) and $\alpha$\,Cen\,B (K1V). The obtained values are in good agreement with observations, which indicate that our simulation of stellar and instrumental noises are realistic and that stars with noise dominated by high and moderate frequency variations exist.
\begin{table*}
	\begin{center}
		\caption{Spectral type, activity level and rms of the residuals, Rms$_{O-C}$, for known planetary systems discovered using HARPS}  \label{tab:0}
		\begin{tabular}{cccccc}
		\hline Star & Spectral type & $\log(R'_{HK})$ & Rms$_{O-C} [ms^{-1}]$ & Reference\\
		\hline
		\hline
		HD69830 & K0V & -4.97 & 0.81 & \citet{Lovis-2006}\\
		HD40307 & K2.5V & -4.99 & 0.85 & \citet{Mayor-2009a}\\
		HD47186 & G5V & -5.01 & 0.91 & \citet{Bouchy-2009}\\
		HD4308 & G5V & -4.93 & 1.3 & \citet{Udry-2006}\\
		HD181720 & G1V & -5.01 & 1.37 & \citet{Santos-2010}\\
		\hline 
		\end{tabular} 
	\end{center}
\end{table*}

\section{Detection limits in the Mass-Period diagram}\label{sec:7}

In order to derive detection limits in terms of planet mass and period accessible with the different studied strategies, we calculate the false alarm probability (FAP) of detection using bootstrap randomization \citep[][]{Endl-2001,Efron-1998}. We first simulate a synthetic RV set containing stellar, instrumental and photon noises as we did in the previous sections. The calendar used (see Sec. \ref{sect:3}) is the same, with 256 nights over a time span of 4 years. Then we carry out 1000 bootstrap randomizations\footnote{Following the use of this well sampled calendar, we choose to calculate the FAP using bootstrap randomization rather than permutation since it is more efficient in this case \citep[][]{Efron-1998}} of these RVs and calculate the corresponding periodograms. For each periodogram, we select the highest peak and construct a distribution of these 1000 highest peaks. The 1\,\% FAP corresponds to the power which is only reached 1\,\% of the time. The second step consist in adding a sinusoidal signal with a given period to the synthetic RVs. We calculate the periodogram of these new RVs and compare the height of the observed peak with the 1\,\% FAP for the given period. We then adjust the semi amplitude of the signal until the power of the peak is equal to the 1\,\% FAP. The obtained semi amplitude corresponds to the detection limit of a null eccentricity planet with a confidence level of 99\,\%. To be conservative, we test 10 different phases and select the highest semi amplitude value.

\begin{figure*}
\centering
\includegraphics[width=8cm]{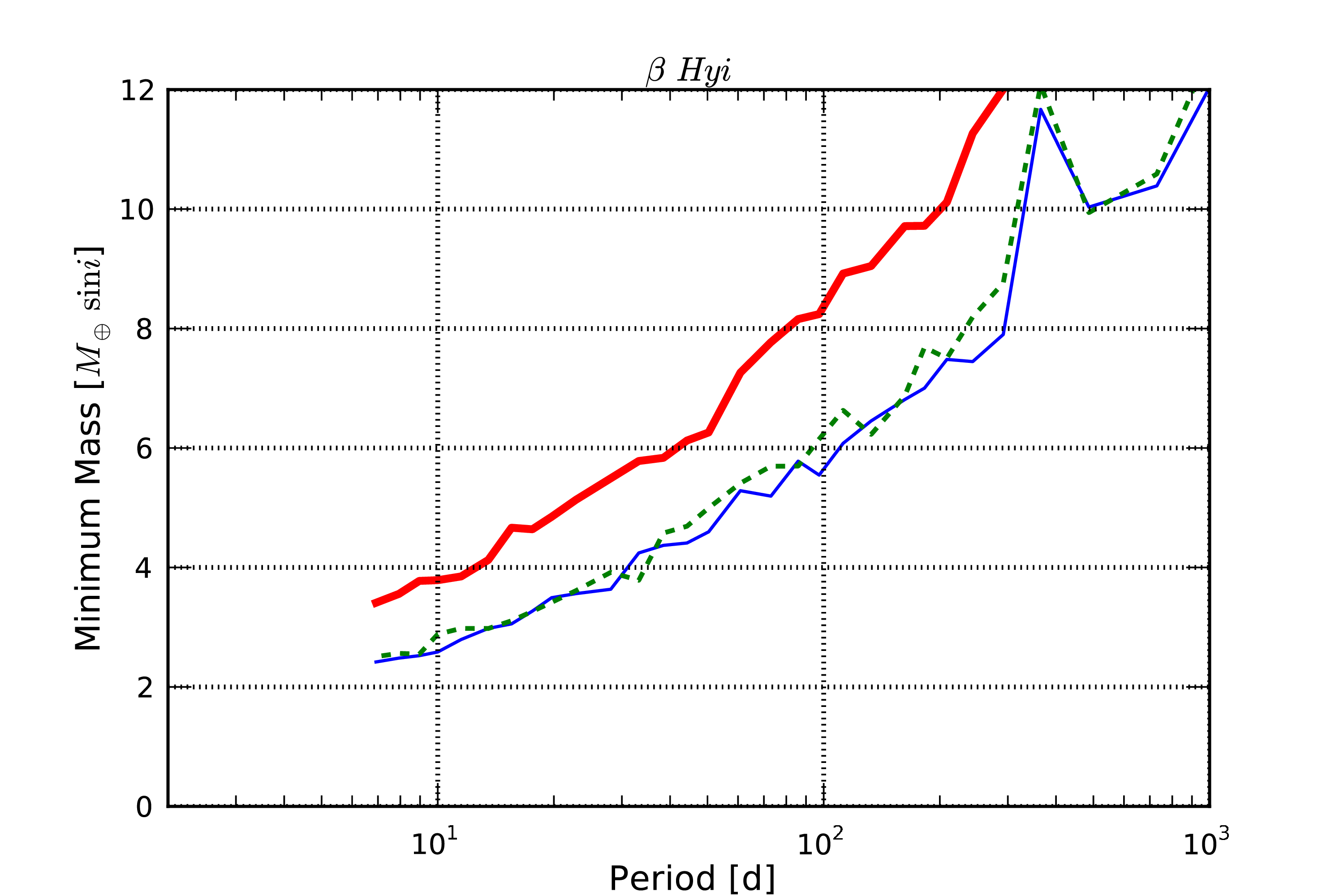}
\includegraphics[width=8cm]{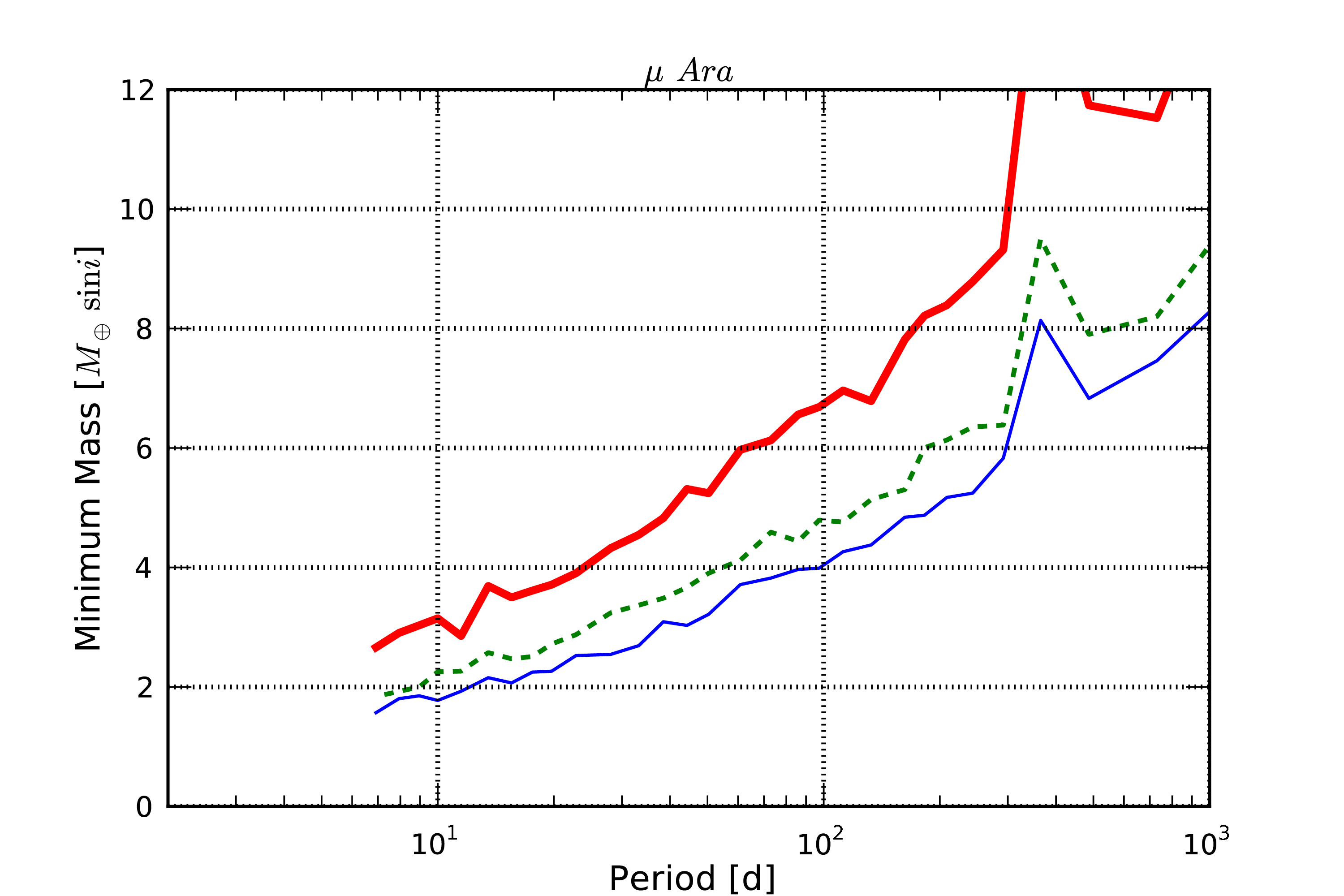}
\includegraphics[width=8cm]{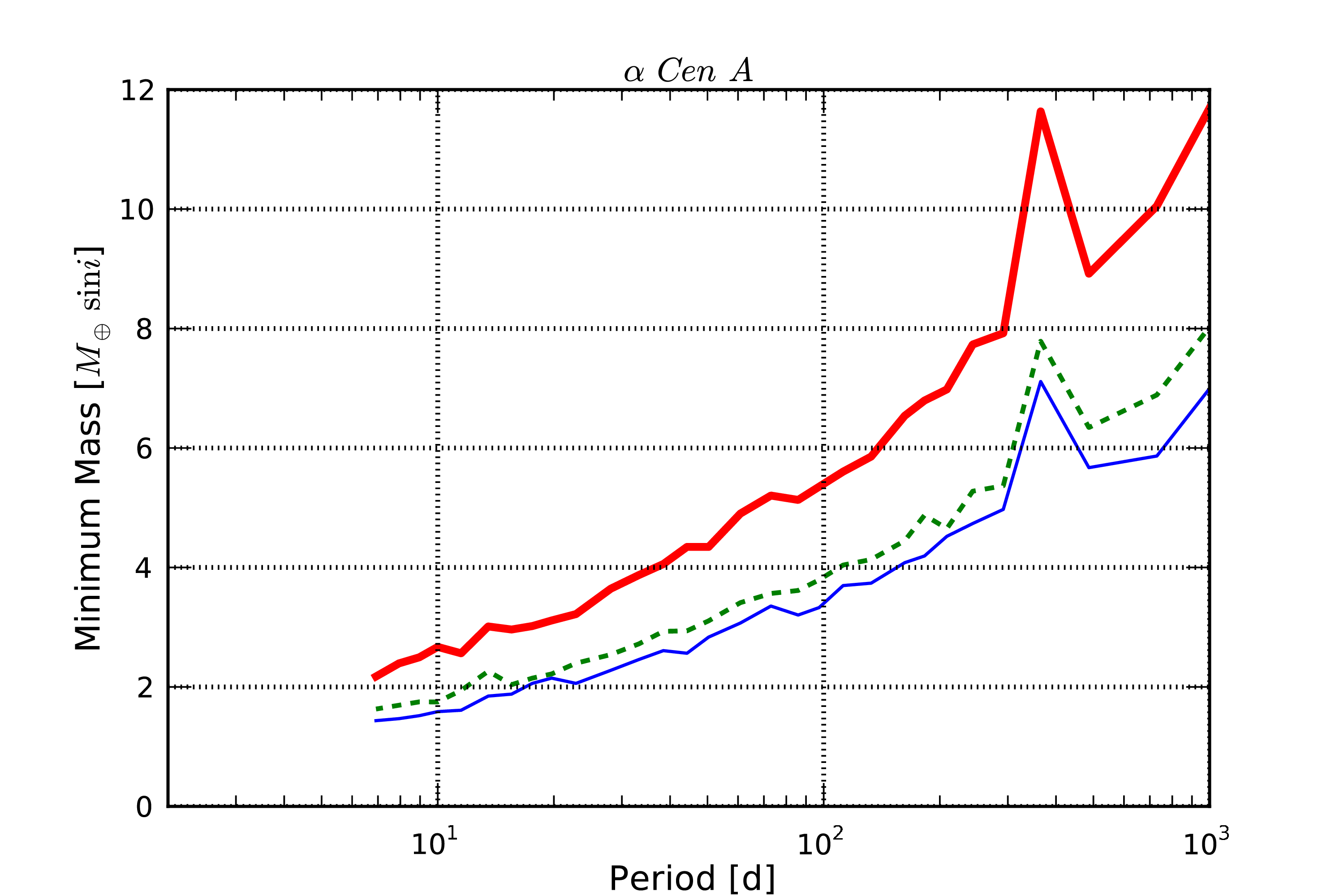}
\includegraphics[width=8cm]{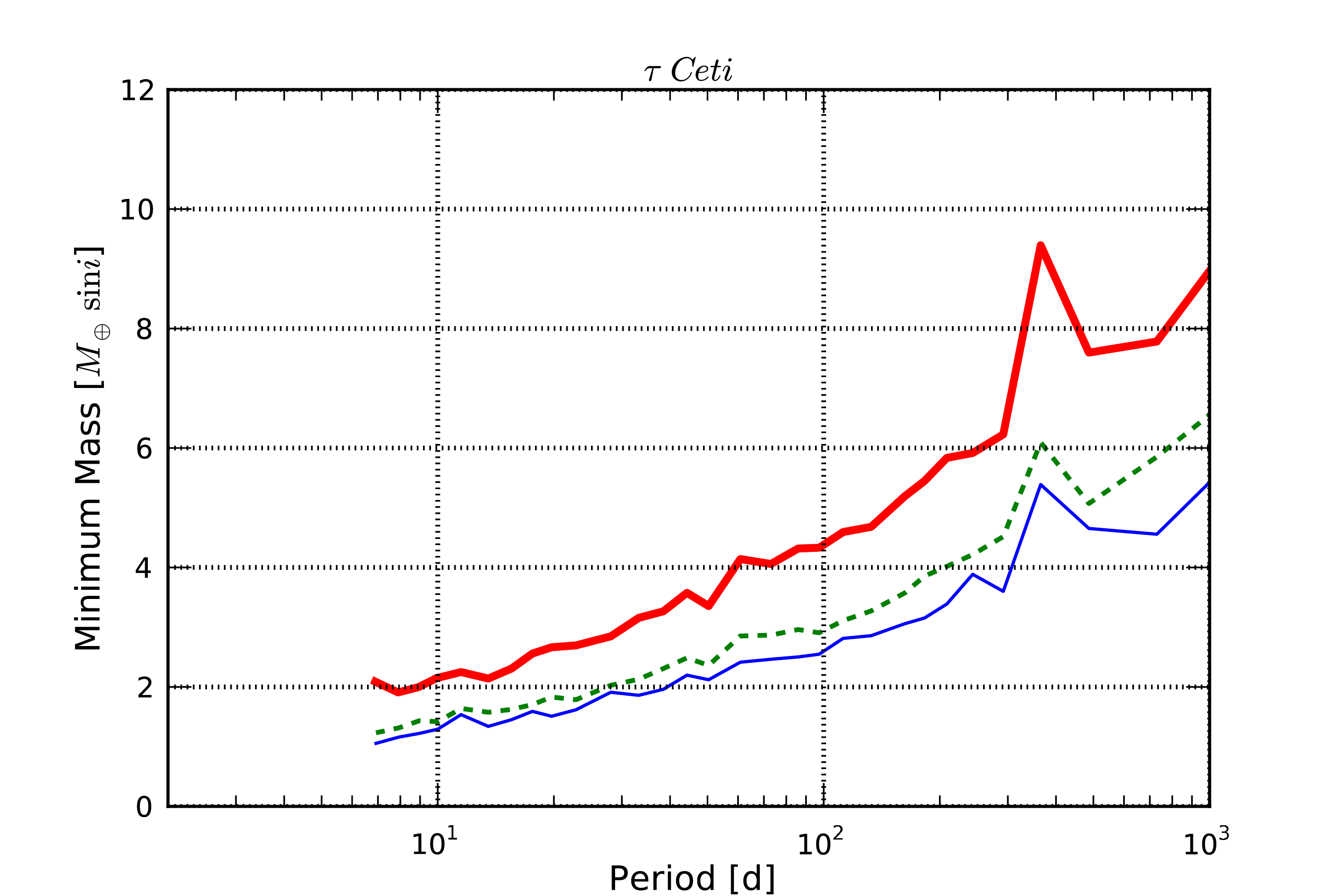}
\includegraphics[width=8cm]{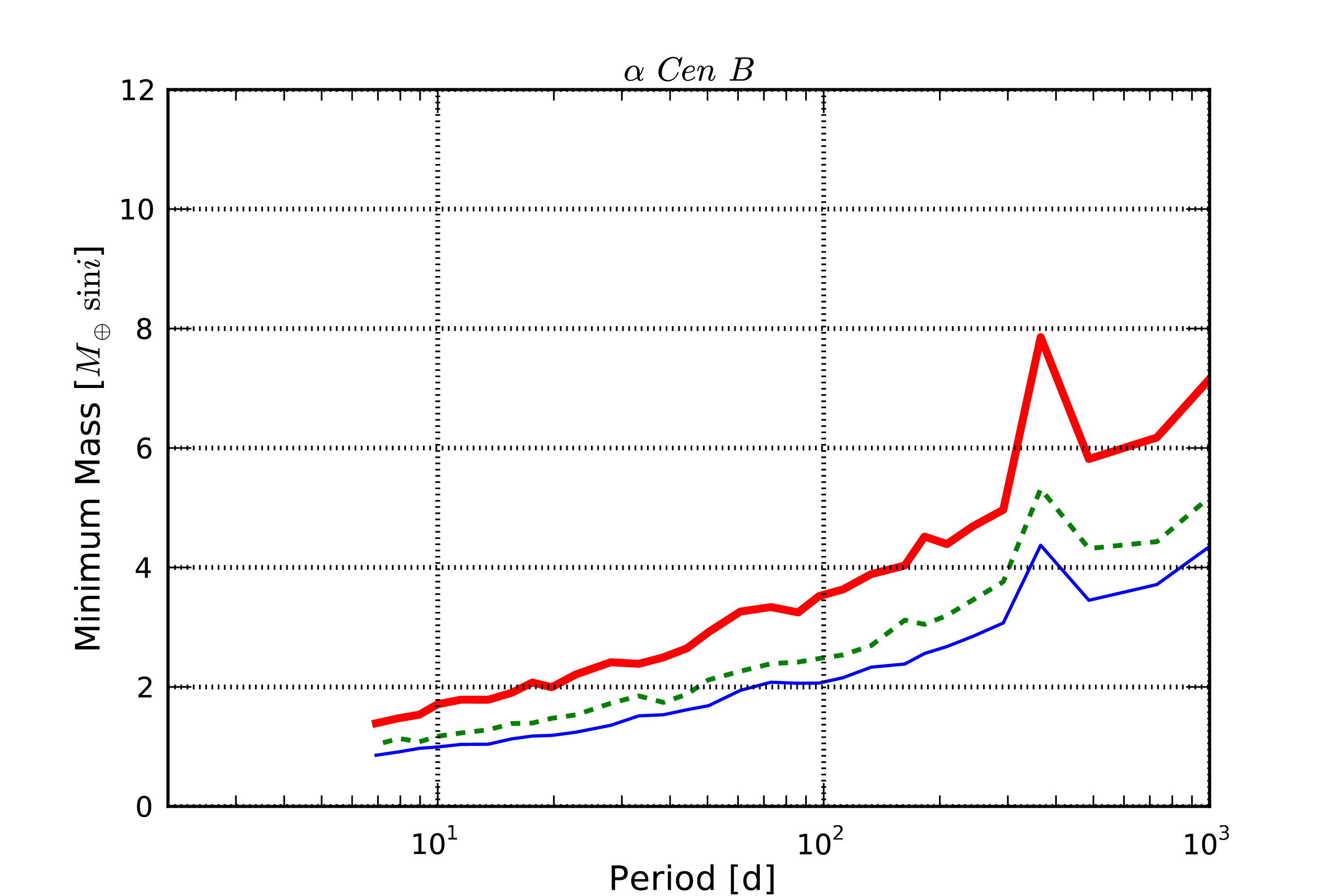}
\caption{Mass period diagrams for our star sample. We can see on each graph, the 3 strategies studied. 1 measurement of 15 minutes per night (continuous thick line),  2 measurements per night of 15 minutes 5 hours apart (dashed line) and 3 measurements per night of 10 minutes 2 hours apart (continuous thin line).}
\label{fig:6}
\end{figure*}

In practice, we simulate 100 RV sets for each strategy, and only the 10 "worst" cases were considered in the end. In Fig.\,\ref{fig:6}, we can see the mass detection limits, with a confidence level of 99\,\%, for the 1N, 2N and 3N strategies using HARPS. 
We note on each graph a peak of the mass detection limits at one year. This can be explain by the 4 months removed each year in the calendar to take into account that stars disappear from the sky. Signals with 1 year of period are not well sampled, which leads to a local raise of the detection limit.

Among the 3 strategies studied, we notice that the use of the 2N and 3N strategies allows us to decrease the level of detection limits by approximatively 2\,$M_{\oplus}$, compared to the high precision strategy normally used on HARPS (1N strategy). This improvement is not negligible, since it reduces mass detection limits by approximately 30\,\%. The difference between the 2N and the 3N strategy is small, but as shown before in Sec \ref{subsect:5.1}, the 3N strategy seems the most efficient one to average out the considered stellar noise.


In Fig\,\ref{fig:6}, stars are ordered from top to bottom and left to right from early spectral type and evolved to late spectral type and not evolved. Following this order, we notice that the level of mass detection limits is going down. Thus, $\alpha$\,Cen\,B, which is a K1 dwarf star, has the smallest mass detection limits in our sample.

If we only look to dwarf stars, $\alpha$\,Cen\,A (G2V), $\tau$\,Ceti (G8V) and $\alpha$\,Cen\,B (K1V), we notice that the level of the mass detection limits is going down when we go towards late spectral types. For example, for 100 days of period, the mass detection limit for the 3N strategy will be 3.5\,M$_{\oplus}$, 2.5\,M$_{\oplus}$ and 2\,M$_{\oplus}$ for the G2V star, the G8V and the K1V, respectively. Thus, late spectral type stars have lower mass detection limits, which can be explained by 2 phenomena. The first one, obvious, is that early K dwarfs will be less massive than G dwarfs. Therefore, for a given mass planet, the RV signal will have a higher variation for late spectral type star. Nevertheless, this effect is not sufficient to explain this large range of mass detection limits\footnote{$\alpha$\,Cen\,B as a masse of 0.90\,$M_{\odot}$, which is slightly smaller than $\alpha$\,Cen\,A, 1.09\,$M_{\odot}$} and the second phenomena, which is the most important one, is that early K dwarfs have a lower level of stellar noise, compared to G dwarfs (see Table\,\ref{tab:2} and Fig\,\ref{fig:4}).

If we now look, in the same spectral type range, to stars with different evolutionary states, we notice also an interesting behavior. $\beta$\,Hyi is clearly a G2 subgiant, $\mu$\,Ara is a G3 star in the transition between the dwarf and the subgiant branch \citep[][]{Soriano-2010}, whereas $\alpha$\,Cen\,A is a G2 dwarf. Studying these 3 different cases, we notice that the level of mass detection limit is higher for evolved stars. For example, the subgiant $\beta$\,Hyi has mass detection limits 1.5 times higher than the dwarf $\alpha$\,Cen\,A.

Following these results on dwarf and sub-giant stars, it seems clear that to find the lowest mass planets using RVs, we should look to early K dwarfs. Such stars present the lowest level of stellar noise and thus, the lowest mass detection limits. Besides this, early K dwarfs are also interesting because their habitable zones, around 200 days, are closer than for early G dwarfs. For stars like $\alpha$\,Cen\,B (K1V) (see Fig. \ref{fig:6}), the simulations predict that we could find planets of 2.5\,M$_{\oplus}$ in the habitable zone, using the 3N strategy.

As already pointed before, due to a short time span of the asteroseismology measurements, stellar activity noise is not fully included in our simulation. Thus, the present limits give  the RV signal that can be reliably detected when activity noise is negligeable. In the presence of real stellar activity noise, the detection limits will raise up. This will be the point in a forthcoming paper.

\section{Detection limits using a real HARPS observational calendar}\label{sect:8}

\begin{figure}[!t]
\centering
\includegraphics[width=8cm]{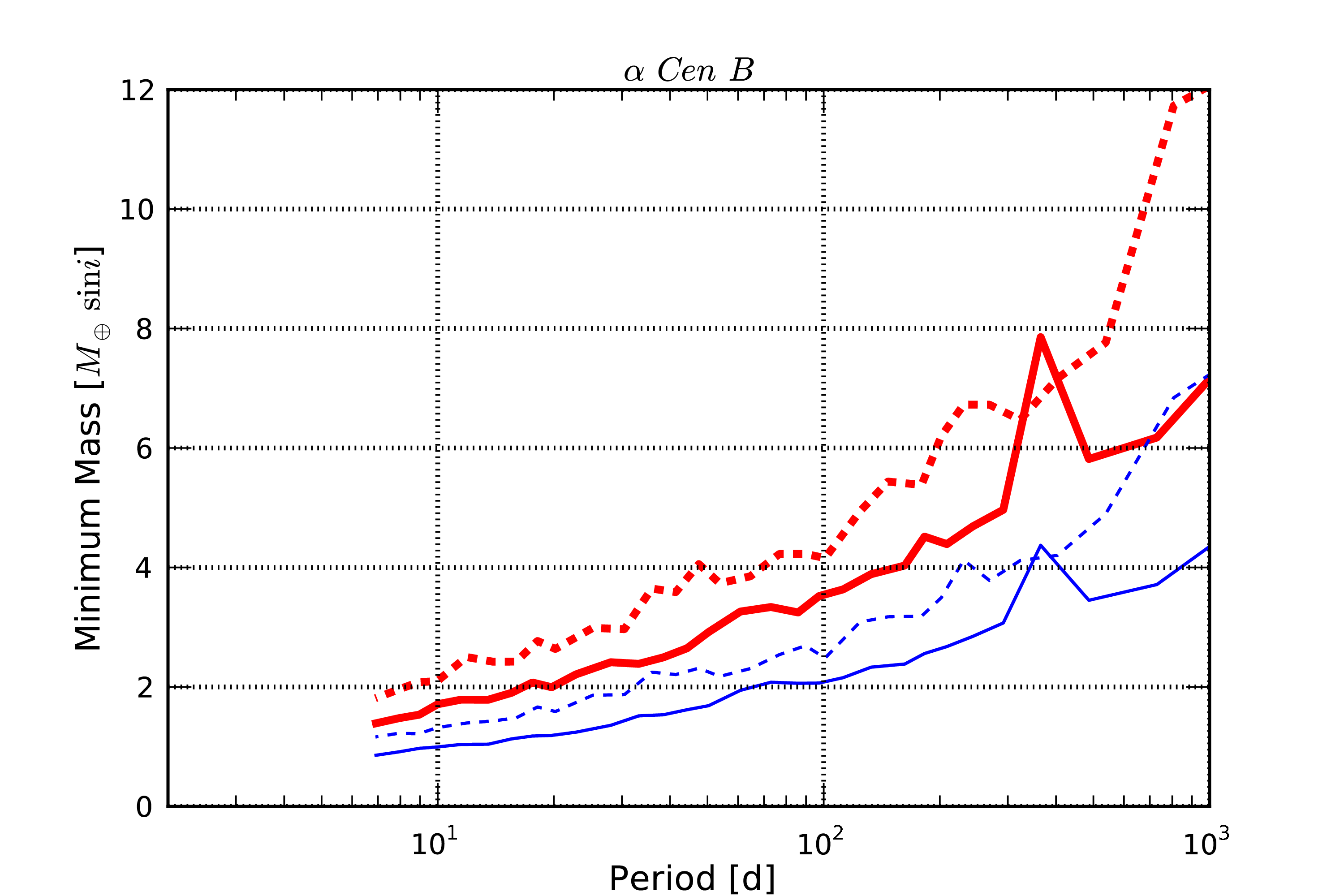}
\caption{Mass period diagrams for $\alpha$\,Cen\,B. Thick and thin lines represent the 1N and the 3N strategy, respectively. Plain lines corresponds to the simulated calendar and dashed lines to the real calendar of HD\,69830.}
\label{fig:7}
\end{figure}

In order to check if the calendar we use is not too idealistic compared to a real one (see section \ref{sect:3}), we compute the detection limits using the real calendar of one of the most followed star using HARPS, HD\,69830. This real calendar regroups a total of 157 nights over an observing span of 1615 days \citep[][Lovis et al. 2010, in preparation]{Lovis-2006}. Figure \ref{fig:7} shows the difference in mass detection limits between the calendar used in this paper and the real one for HD\,69830.
At 100 days of period, the detection limit for the 2 calendar are 2 and 2.5\,M$_{\oplus}$, respectively. This small increase of 25\,\% is only due to the total number of night present in each calendar. The semi amplitude in RV goes like the square root of the total number of measurement. Therefore, passing from 256 nights to 157 will increase the detection limit by 28\,\%, very close to what gives the simulation. Applying the 3N strategy on a K dwarf, such as HD\,69830, should lead to detection of 3\,M$_{\oplus}$ planets in their habitable zone. Thus, by using a real HARPS observational calendar, we can reach similar detection limits. A real observational calendar depends not only on bad weather and instrumental problems, but also to the following time allocated for each star. The present strategy on HARPS is to follow a lot of targets, which reduce the number of measurement per star. In searching for the smallest planets, just a very small sample of stars should be selected. This would allow us to increase the number of measurement compared to a real observational HARPS calendar. It is why the simulations, done in the previous section, use 256 nights of measurements on 4 years rather than 157.

\section{Concluding remarks}

The present high-precision HARPS observational strategy (1 measurement of 15 minutes per night, 1N strategy in the paper) manages well to reduce stellar oscillation noises, but is not optimized to average out granulation phenomena (granulation, mesogranulation and supergranulation). These latter sources of noise perturb radial velocity measurements on time scales up to 1.5 days. Increasing the number of measurements per night allows us to reduce the effect of granulation phenomena, after binning, and consequently improve the planetary detection limits. The best tested observational strategy, 3 measurements per night each 2 hours, with an exposure time of 10 minutes each (3N strategy in the paper), gives detection limits on average 30\,\% better than the present HARPS high-resolution strategy. This improvement is really due to the 3 measurements per night and not to the total observational time (which is doubled compared to the present HARPS observational strategy). Indeed, we have shown that 1 measurement per night of 30 minutes is not efficient.

Our work also suggests a trend between the level of the stellar noise considered in this study and the spectral type for dwarf stars. Early K dwarfs have a total noise level lower than early G dwarfs. A trend between evolved and non evolved stars is also revealed, non evolved stars showing a lower level of stellar noise. Therefore, early K dwarfs seems to have the lowest level of stellar noise of our sample and consequently the lowest planet mass detection limits. In addition, the habitable zone is closest in early K than in early G dwarfs, making early K dwarfs the most promising targets to search for very low mass planets.

Following the conclusion of the last 2 paragraphs, we applied the 3 measurements per night strategy on $\alpha$\,Cen\,B (K1V). For the habitable zone of this planet, approximatively 200 days, our simulation shows that we would be able to detect planets of 3 M$_{\oplus}$ using an existing HARPS observational calendar. Therefore, granulation phenomena and oscillation modes will not prevent us of finding Earth like planets in habitable regions.

However, there are 2 different limitations to our present study. First of all, due to only short time span measurements, we do not take properly into account stellar activity noise sources, which can be important. The detection limits we obtain are only valid for stars without significant activity related phenomena, such as spots and plages. We have shown that such stars exist and therefore activity will not be a problem for some targets. A second limitation could come from the fact that we only simulate the presence of one planet per star. The majority of small planets discovered nowadays are in multiple systems, which complicate the analysis. More data is simply needed in these cases to allow us to fit all the free parameters. 

Next generation spectrographs, such as ESPRESSO (http://espresso.astro.up.pt/) and CODEX \citep[e.g][]{Pasquini-2008}, will reach better levels of precision and stability, most likely leading to a reduction of detection limits near 1 M$_{\oplus}$.

\begin{acknowledgements}

We would like to thank Tim Bedding for providing us the power spectrum and the window function of $\beta$\,Hyi.  N.C. Santos would like to thank the support by the European Research Council/European Community under the FP7 through a Starting Grant, as well as the support from Funda\c{c}\~ao para a Ci\^encia e a Tecnologia (FCT), Portugal, through programme Ci\^encia\,2007. We would also like to acknowledge support from FCT in the form of grants reference PTDC/CTE-AST/098528/2008 and PTDC/CTE-AST/098604/2008. Finally, this work was supported by the European Helio- and Asteroseismology Network (HELAS), a major international collaboration funded by the European Commission's Sixth Framework Program (grant : FP6-2004-Infrastructures-5-026183).

\end{acknowledgements}

\bibliographystyle{aa}
\bibliography{dumusque_bibliography}

\end{document}